\documentclass[11pt]{article}

\usepackage{amsfonts,amsmath,amssymb,bm,graphicx,hyperref,a4wide,cite,mathabx,subfigure}

\begin{document}

\title{\textbf{Spin effects in neutrino gravitational scattering}}

\author{Maxim Dvornikov\thanks{maxdvo@izmiran.ru} 
\\
\small{\ Pushkov Institute of Terrestrial Magnetism, Ionosphere} \\
\small{and Radiowave Propagation (IZMIRAN),} \\
\small{108840 Troitsk, Moscow, Russia}}

\date{}

\maketitle

\begin{abstract}
We study spin oscillations of neutrinos gravitationally scattered
off a nonrotating black hole (BH). We derive the transition and survival
probabilities of spin oscillations in quadratures when neutrinos interacts
with BH only. The dependence of the probabilities on the impact parameter
is analyzed. Then, we obtain the effective Schr\"{o}dinger equation for
neutrino spin oscillations in neutrino scattering off BH surrounded
by background matter. This equation is solved numerically in the case
of a supermassive BH with a realistic accretion disk. We find that
the observed neutrino fluxes can be reduced almost 20\% because of
spin oscillations when neutrinos experience gravitational scattering.
The neutrino interaction with an accretion disk results in the additional
asymmetry in the intensities of outgoing fluxes depending on the neutrino
trajectory.
\end{abstract}

\section{Introduction}

The experimental observation of neutrino oscillations, reported, e.g.,
in Ref.~\cite{Ago18}, confirmed that neutrinos are massive particles
having nonzero mixing between different flavors. Among various types
of neutrino oscillations, we distinguish neutrino spin oscillations~\cite{GiuStu15},
which are the transitions between different helicity states within
one neutrino type. If a left polarized neutrino changes its polarization,
it cannot be observed since right neutrinos are sterile in the standard
model. It will result in the effective reduction of the initial neutrino
flux.

It is known that external backgrounds, e.g., the neutrino interaction
with matter~\cite{MalSmi16}, can modify the process of neutrino
oscillations. The gravitational interaction was found in Ref.~\cite{AhlBur96}
to influence flavor oscillations of neutrinos. Neutrino spin oscillations
in various external fields in curved spacetime were studied in Refs.~\cite{Dvo06,Dvo13,Dvo19}
We considered both static metrics and a time dependent backgrounds,
like a gravitational wave. Note that the evolution of the fermion
spin in curved spacetime was analyzed in Refs.~\cite{ObuSilTer09,ObuSilTer17}
using both quasiclassical and quantum approaches.

The gravity induced neutrino spin oscillations, studied in Refs.~\cite{Dvo06,Dvo13,SorZil07,AlaNod15,Cha15},
were analyzed for neutrinos orbiting a massive object, e.g., a black
hole (BH). However, in this situation, even if neutrino spin oscillations
can be quite intense, it is rather difficult to understand what kind
of observational effects one can expect since a particle is gravitationally
captured by BH, or a neutrino falls to the BH surface. That is why
it is interesting to study spin effects, or neutrino spin oscillations,
e.g., in the neutrino gravitational scattering, when one can control
the helicities of both incoming and outgoing particles.

This research is inspired by the recent observation of the shadow
of a supermassive BH (SMBH)~\cite{Aki19}, which provides the unique
test of the general relativity in the strong field limit. A bright
ring around a BH shadow is formed by photons, which are emitted by
an accretion disk and then experience strong lensing in the gravitational
field of BH~\cite{GraHolWal19}. However, besides photons, a significant
flux of neutrinos was found in Ref.~\cite{CabMcLSur12} to be emitted
by an accretion disk. These particles are subject to neutrino oscillations.
In this work, we shall examine how a strong gravitational field of
BH and the neutrino interaction with an accretion disk can modify
the helicity of scattered particles.

The neutrino gravitational scattering was studied recently~\cite{Cor15},
mainly in connection with the determination of the BH shadow produced
by these particles~\cite{StuSch19}. In our work, we shall focus
on the analysis of spin oscillations in the neutrino gravitational
scattering, which effectively reduce the flux of neutrinos measured
in a detector.

Photons, which form the ring around the BH shadow, interact both with
its gravitational field and with plasma which surrounds BH. The interaction
with plasma can modify the size and the form of the BH shadow (see
Ref.~\cite{CunHer18} for a review). In the present work, we shall
study how the neutrino interaction with background matter, e.g., with
an accretion disk, can influence the observed flux of gravitationally
scattered neutrinos.


In this our work, we continue our studies of neutrino spin oscillations
in Refs.~\cite{Dvo06,Dvo13,Dvo19}. We start in Sec.~\ref{sec:GRAV}
with the analysis of the neutrino spin evolution when a particle gravitationally
scatters off a nonrotating BH. We find the expressions in quadratures
for the transition and survival probabilities for ultrarelativistic
neutrinos and analyze them for different impact parameters. Then,
in Sec.~\ref{sec:MATT}, we formulate the effective Schr\"{o}dinger equation
for neutrino spin oscillations in the scattering off BH surrounded
by background matter. We study astrophysical applications in Sec.~\ref{sec:APPL}.
In particular, we consider the effect of spin oscillations on the
measured neutrino fluxes when particles scatter off SMBH with a realistic
accretion disk. Finally, in Sec.~\ref{sec:DISC}, we discuss our
results. We remind how a scalar particle
moves in the Schwarzschild metric in Appendix~\ref{sec:PARTM}.

\section{Neutrino spin evolution in scattering off BH\label{sec:GRAV}}

In this section, we study how the spin of a neutrino evolves when
a particle scatters off a Schwarzschild BH. We solve the spin evolution
equation in quadratures and analyze the solution for ultrarelativistic
neutrinos. The transition and survival probabilities for neutrino
spin oscillations are derived.%

We study the neutrino motion in the vicinity of a nonrotating BH.
Using the spherical coordinates $(r,\theta,\phi)$, the interval in
this case has the form~\cite[p.~284]{LanLif71},
\begin{equation}
\mathrm{d}\tau^{2}=A^{2}\mathrm{d}t^{2}-A^{-2}\mathrm{d}r^{2}-r^{2}(\mathrm{d}\theta^{2}+\sin^{2}\theta\mathrm{d}\phi^{2}),\label{eq:intschw}
\end{equation}
where $A=\sqrt{1-r_{g}/r}$, $r_{g}=2M$ is the gravitational radius,
and $M$ is the BH mass. Since the Schwarzschild metric in Eq.~(\ref{eq:intschw})
is spherically symmetric, we can take that a neutrino moves in the
equatorial plane with $\theta=\pi/2$, i.e. $\mathrm{d}\theta=0$.

In Refs.~\cite{Dvo06,Dvo13}, we found that the neutrino invariant
spin $\bm{\zeta}$, defined in a locally Minkowskian frame, evolves
as
\begin{equation}
\frac{\mathrm{d}\bm{\zeta}}{\mathrm{d}t}=\frac{2}{\gamma}(\bm{\zeta}\times\bm{\Omega}_{g}),\label{eq:spinevgen}
\end{equation}
where $\gamma=\mathrm{d}t/\mathrm{d}\tau$. If a neutrino interacts
with a Schwarzschild BH, the vector $\bm{\Omega}_{g}$ in Eq.~(\ref{eq:spinevgen})
has only one nonzero component~\cite{Dvo06},
\begin{equation}
\bm{\Omega}_{g}=(0,\Omega_{2},0),\quad\Omega_{2}=\frac{1}{2}\frac{\mathrm{d}\phi}{\mathrm{d}t}\left(-A+\frac{\gamma}{\left(1+\gamma A\right)}\frac{r_{g}}{2r}\right),\label{eq:Omega2}
\end{equation}
where $\mathrm{d}\phi/\mathrm{d}t=LA^{2}/Er^{2}$ is the angular velocity,
which can be obtained using Eq.~(\ref{eq:eqmtr}), $L$ is the conserved
angular momentum of a neutrino, and $E$ is the neutrino energy. The
parameter $\gamma$ in Eqs.~(\ref{eq:spinevgen}) and~(\ref{eq:Omega2})
has the form, $\gamma=E/mA^{2}$.

We are interested in neutrino spin oscillations, i.e. in the change
of the neutrino helicity, $h=(\bm{\zeta}\mathbf{u})/|\mathbf{u}|$,
where $\mathbf{u}$ is the spatial part of the neutrino four velocity
in the locally Minkowskian frame. Therefore, besides the study of
the neutrino spin in Eq.~(\ref{eq:spinevgen}), we should account
for the evolution of $\mathbf{u}$.

The expression for $\mathbf{u}$ in the Schwarzschild metric has the
form~\cite{Dvo06},
\begin{align}
\mathbf{u}= & \left(\frac{\mathrm{d}r}{\mathrm{d}\tau}A^{-1},0,r\frac{\mathrm{d}\phi}{\mathrm{d}\tau}\right)\nonumber \\
 & =\left(\pm\frac{1}{m}\left[E^{2}-m^{2}A^{2}\left(1+\frac{L^{2}}{m^{2}r^{2}}\right)\right]^{1/2},0,\frac{L}{mr}\right),\label{eq:uexpl}
\end{align}
where the signs $\pm$ stay for outgoing and incoming neutrinos respectively
(see Eq.~(\ref{eq:eqmtr})). At $r\to\infty$, $\mathbf{u}\to\mathbf{u}_{\pm\infty}=\left(\pm\left[E^{2}-m^{2}\right]^{1/2}/m,0,0\right)$,
i.e. the asymptotic neutrino motion happens along the first axis in
the locally Minkowskian frame. In this frame, an incoming neutrino
propagates oppositely the first axis. An outgoing particle moves along
this axis.

Since only $\Omega_{2}\neq0$, the nonzero neutrino spin components
are $\zeta_{1,3}\neq0$, and $\zeta_{2}=0$. It is convenient to represent
\begin{equation}
\zeta_{1}=\cos\alpha,\quad\zeta_{3}=\sin\alpha,\label{eq:zeta13alpha}
\end{equation}
where $\alpha$ is the rotation angle of the spin from its initial
direction.

Now we have to specify the initial condition for Eq.~(\ref{eq:spinevgen}).
We suppose that, initially, at $r\to\infty$ and $\phi\to0$, an incoming
neutrino is left polarized, i.e. the helicity is negative, $h_{-\infty}=(\bm{\zeta}_{-\infty}\mathbf{u}_{-\infty})/|\mathbf{u}_{-\infty}|=-1$.
Accounting for the expression for $\mathbf{u}_{-\infty}$ above, we
get that $\zeta_{-\infty1}=1$ and $\zeta_{-\infty3}=0$, or $\alpha_{-\infty}=0$
in Eq.~(\ref{eq:zeta13alpha}).

The helicity of an outgoing neutrino has the form, $h_{+\infty}=(\bm{\zeta}_{+\infty}\mathbf{u}_{+\infty})/|\mathbf{u}_{+\infty}|$,
where $\bm{\zeta}_{+\infty}=(\cos\alpha_{+\infty},0,\sin\alpha_{+\infty})$
and $\mathbf{u}_{+\infty}$ is given above. Using Eq.~(\ref{eq:zeta13alpha}),
we get that $h_{+\infty}=\cos\alpha_{+\infty}$. The transition $P_{\mathrm{LR}}$
and survival $P_{\mathrm{LL}}$ probabilities for neutrino spin oscillations
are
\begin{equation}
P=\frac{1}{2}(1\pm h_{+\infty}),\label{eq:Pgen}
\end{equation}
where the upper sign stays for $P_{\mathrm{LR}}$ and the lower one
for $P_{\mathrm{LL}}$.

The angle $\alpha$ corresponds to the spin projection on the $x$-axis
in the neutrino rest frame. This projection is in $m/E$ times shorter
for a nonmoving observer, which measures the neutrino polarization,
because of the Lorentz contraction. It means that the observed angle
should be rescaled by the factor $E/m$: $\alpha\to\alpha E/m$. It is
equivalent to the replacement of $\Omega_{2}$: $\Omega_{2}\to\Omega_{2}E/m$.

Now we should find $\alpha_{+\infty}$. Using Eqs.~(\ref{eq:spinevgen}),
(\ref{eq:Omega2}), (\ref{eq:zeta13alpha}) and~(\ref{eq:eqmtr}),
we get that the angle $\alpha$ obeys the equation,
\begin{align}
\frac{\mathrm{d}\alpha}{\mathrm{d}r}= & F,\quad F(r)=\pm\frac{AL}{mr^{2}}\frac{\frac{E}{m}\left(\frac{3r_{g}}{2r}-1\right)-A^{3}}{\frac{E}{m}+A}\left[\frac{E^{2}}{m^{2}}-A^{2}\left(1+\frac{L^{2}}{m^{2}r^{2}}\right)\right]^{-1/2},\label{eq:F}
\end{align}
where the signs $\pm$ stay for outgoing and incoming neutrinos. Then
we should account for the initial condition $\alpha_{-\infty}=0$
and the fact that $\alpha_{+\infty}$ is twice the angle corresponding
to the minimal distance between a neutrino and BH. We express the
final result for ultrarelativistic neutrinos, with $E\gg m$, as
\begin{equation}
\alpha_{+\infty}=y\int_{x_{m}}^{\infty}\frac{\mathrm{d}x}{x^{2}}\frac{(3-2x)\sqrt{x-1}}{\sqrt{x^{3}-y^{2}(x-1)}},\label{eq:aobslim}
\end{equation}
where $y=b/r_{g}$, $b=L/E$ is the impact parameter, and $x_{m}$
is the maximal root of the equation 
\begin{equation}
x^{3}-y^{2}(x-1)=0.\label{eq:eqtosolve}
\end{equation}
Note that $y>y_{0}=3\sqrt{3}/2$ for a neutrino not to fall to BH
(see Appendix~\ref{sec:PARTM}).

The expression for roots $x_{1,2,3}$ of Eq.~(\ref{eq:eqtosolve})
for the arbitrary $y$ has the form,
\begin{equation}
x_{k}=\frac{2y}{\sqrt{3}}\cos\left[\frac{1}{3}\arccos\left(-\frac{3\sqrt{3}}{2y}\right)-\frac{2\pi}{3}(k-1)\right],\quad k=1,2,3,\label{eq:roots}
\end{equation}
where $x_{1}\equiv x_{m}$ is the maximal root. First, let us analyze
Eq.~(\ref{eq:aobslim}) in the case $y\gg y_{0}$. Using Eq.~(\ref{eq:roots})
and keeping only the leading terms, we get that the roots have the
form, $x_{1}=y-\tfrac{1}{2}-\tfrac{3}{8y}+\mathcal{O}(y^{-3})$, $x_{2}=1+\mathcal{O}(y^{-4})$,
and $x_{3}=-y-\tfrac{1}{2}+\tfrac{3}{8y}+\mathcal{O}(y^{-3})$. In
this case, we get that
\begin{equation}
\alpha_{+\infty}=8y\int_{a}^{\infty}\frac{\mathrm{d}x}{(2x-1)^{2}}\frac{(1-x)}{\sqrt{x^{2}-a^{2}}}\approx-\pi-\frac{\pi}{4y^{2}},\label{eq:abigy}
\end{equation}
where $a=y-\tfrac{3}{8y}$. The transition and survival probabilities
in Eq.~(\ref{eq:Pgen}) take the form,
\begin{equation}
P_{\mathrm{LR}}=\frac{1}{2}\left[1-\cos\frac{\pi}{4y^{2}}\right]\approx\frac{\pi^{2}}{64y^{4}},\quad P_{\mathrm{LL}}=\frac{1}{2}\left[1+\cos\frac{\pi}{4y^{2}}\right]\approx1-\frac{\pi^{2}}{64y^{4}}.\label{eq:Plim}
\end{equation}
One can see that $P_{\mathrm{LR}}\to0$ (and $P_{\mathrm{LL}}\to1$)
if $y\gg y_{0}$. This is expected since, at $y\gg y_{0}$, a neutrino
propagates far away from BH. The gravitational interaction, which
causes the spin flip, is weak. Thus the neutrino polarization is practically
unchanged.

Now we discuss the situation when $y\to y_{0}$. Then, Eq.~(\ref{eq:eqtosolve})
has the following roots: $x_{1}=x_{2}=3/2$ and $x_{3}=-3$. The spin
rotation angle takes the value
\begin{equation}
\alpha_{+\infty}=-3\sqrt{3}\int_{3/2}^{\infty}\frac{\mathrm{d}x}{x^{2}}\frac{\sqrt{x-1}}{\sqrt{x+3}}=-\frac{2\pi}{3},
\end{equation}
which is finite even if a neutrino asymptotically approaches BH. The
corresponding probabilities are $P_{\mathrm{LR}}=0.25$ and $P_{\mathrm{LL}}=0.75$
for such neutrinos. We shall present the transition and survival probabilities
for arbitrary $y$ in Sec.~(\ref{sec:APPL}), when we study some
possible astrophysical applications.

\section{Neutrino gravitational scattering accounting for the matter interaction\label{sec:MATT} }

In this section, we formulate the neutrino spin evolution equations
in background matter under the influence of a gravitational field
when a neutrino scatters off BH. Then, we derive the effective Schr\"{o}dinger
equation for scattered neutrinos.

Using the forward scattering approximation, one gets that the neutrino
interaction with background matter is described by the following effective
Lagrangian in Minkowsky spacetime~\cite{MohPal04}:
\begin{equation}
\mathcal{L}_{m}=-\sqrt{2}G_{\mathrm{F}}\bar{\nu}\gamma^{\mu}(1-\gamma^{5})\nu G_{\mu},\label{eq:Largmat}
\end{equation}
where $\nu$ is the neutrino bispinor, $\gamma^{\mu}$ and $\gamma^{5}$
are the Dirac matrices, and $G_{\mathrm{F}}=1.17\times10^{-5}\,\text{GeV}^{-2}$
is the Fermi constant. The four vector $G^{\mu}$ is the linear combination
of the hydrodynamic currents and polarizations of background fermions.
It depends on the chemical composition of matter and the type of the
neutrino. The explicit form of $G^{\mu}$ can be found in Ref.~\cite{DvoStu02}.

Basing on Eq.~\ref{eq:Largmat}, the influence of the neutrino interaction
with background matter on its spin evolution in curved spacetime was
studied in Refs.~\cite{Dvo13,Dvo19}. It results in the appearance
of the additional components of the vector $\bm{\Omega}_{g}$ in Eq.~(\ref{eq:Omega2}):
$\bm{\Omega}_{g}\to\bm{\Omega}=\bm{\Omega}_{g}+\bm{\Omega}_{m}$.
If we study the neutrino interaction with nonmoving and unpolarized
background fermions in curved spacetime, the vector $\bm{\Omega}_{m}$
has the form,
\begin{equation}
\bm{\Omega}_{m}=\frac{G_{\mathrm{F}}}{\sqrt{2}}\frac{g^{0}}{\gamma}\mathbf{u}=\frac{G_{F}}{\sqrt{2}}n_{\mathrm{eff}}\frac{1}{\gamma}\left(\frac{\mathrm{d}r}{\mathrm{d}\tau},0,Ar\frac{\mathrm{d}\phi}{\mathrm{d}\tau}\right),\label{eq:Omegam}
\end{equation}
where $g^{0}=e_{\,\mu}^{0}G^{\mu}=AG^{0}$, $e_{\,\mu}^{0}=(A,0,0,0)$
is the vierbein vector in the Schwarzschild metric (see Ref.~\cite{Dvo06}),
and $G^{0}=n_{\mathrm{eff}}$ is the invariant effective density of
background matter. We use Eq.~(\ref{eq:uexpl}) to derive Eq.~(\ref{eq:Omegam}).

If we study spin oscillations of electron neutrinos in the electrically
neutral hydrogen plasma then $n_{\mathrm{eff}}=n_{e}$, where $n_{e}$
is the electron number density. The expressions for $n_{\mathrm{eff}}$
for other neutrino oscillations channels and various types background
fermions can be found in Ref.~\cite{DvoStu02}.

Instead of dealing with Eq.~(\ref{eq:spinevgen}) for the spin precession
it is convenient to study the neutrino polarization density matrix,
$\rho=\tfrac{1}{2}[1+(\bm{\sigma\zeta})]$, which obeys the equation,
$\mathrm{i}\dot{\rho}=[H,\rho]$, where $H=-(\bm{\sigma\Omega})$
and $\bm{\Omega}$ includes both the gravity and matter contributions
in Eqs.~(\ref{eq:Omega2}) and~(\ref{eq:Omegam}).

Since the Liouville\textendash von Neumann equation for the density
matrix is rather complicated for the analysis, we can use the Schr\"{o}dinger
equation, $\mathrm{i}\dot{\psi}=H\psi$. As we mentioned in Sec.~(\ref{sec:GRAV}),
neutrinos move along the first axis in the locally Minkowskian frame
at $r\to\infty$. Hence, it is convenient to use this axis for the
spin quantization. It mean that we should transform the Hamiltonian
$H\to U_{2}HU_{2}^{\dagger}$, where $U_{2}=\exp(i\pi\sigma_{2}/4)$.
This procedure brings the meaning to the effective wave function $\psi$.

As in Eq.~(\ref{eq:F}), it is convenient to rewrite the Schr\"{o}dinger
equation using the radial coordinate $r$,
\begin{equation}
\mathrm{i}\frac{\mathrm{d}\psi}{\mathrm{d}r}=H_{r}\psi,\quad H_{r}=-U_{2}(\bm{\sigma\Omega}_{r})U_{2}^{\dagger},\label{eq:Schr}
\end{equation}
where
\begin{equation}
\bm{\Omega}_{r}=\frac{\mathrm{d}t}{\mathrm{d}r}\bm{\Omega}=\left(\frac{G_{F}}{\sqrt{2}}n_{\mathrm{eff}},\frac{F}{2},Ar\frac{\mathrm{d}\phi}{\mathrm{d}r}\frac{G_{F}}{\sqrt{2}}n_{\mathrm{eff}}\right).\label{eq:Omegar}
\end{equation}
Here $F$ is given in Eq.~(\ref{eq:F}).%

Equation~(\ref{eq:Schr}) should be supplied with the initial condition
$\psi_{-\infty}^{\mathrm{T}}=(1,0)$, which means that all incoming
neutrinos are left polarized. Since the neutrino velocity changes
the direction at $t\to+\infty$, the transition probability reads
$P_{\mathrm{LR}}=|\psi_{+\infty}^{(1)}|^{2}$, and, correspondingly,
the survival probability is $P_{\mathrm{LL}}=|\psi_{+\infty}^{(2)}|^{2}$,
where $\psi_{+\infty}^{\mathrm{T}}=(\psi_{+\infty}^{(1)},\psi_{+\infty}^{(2)})$
is the asymptotic solution of Eq.~(\ref{eq:Schr}).

The solution of Eqs.~(\ref{eq:Schr}) and~(\ref{eq:Omegar}) can
be found only numerically because of the nontrivial dependence of
$\bm{\Omega}_{r}$ on $r$. Moreover, in Sec.~\ref{sec:APPL}, we
discuss the situation when $n_{\mathrm{eff}}=n_{\mathrm{eff}}(r)$,
which make the analysis more complicated.

We also mention, that we cannot integrate Eqs.~(\ref{eq:Schr}) and~(\ref{eq:Omegar})
to the turn point $r_{m}$ and then automatically reconstruct $\psi_{+\infty}$,
as we made in Sec.~(\ref{sec:GRAV}) to find $\alpha_{+\infty}$.
In the presence of the background matter, the dynamics of the neutrino
polarization is nonabelian. Moreover, the term $\tfrac{\mathrm{d}\phi}{\mathrm{d}r}$
in $\bm{\Omega}_{r}$ changes the sign at $r_{m}$ (see Eq.~(\ref{eq:eqmtr})).
Thus, to obtain $\psi_{+\infty}$, one should integrate Eqs.~(\ref{eq:Schr})
and~(\ref{eq:Omegar}), first, in the interval $+\infty>r>r_{m}$
and, then, for $r_{m}<r<+\infty$, with the solutions being stitched
at $r_{m}$. This fact significantly reduces the accuracy of the numerical
simulation compared to Sec.~(\ref{sec:GRAV}).

\section{Astrophysical applications\label{sec:APPL}}

In this section, we present the numerical solutions of Eqs.~(\ref{eq:Schr})
and~(\ref{eq:Omegar}) for the neutrino scattering off SMBH surrounded
by an accretion disk. We discuss different orientations of neutrino
trajectories with respect to the disk plane. Measurable neutrino fluxes
are obtained.

First we notice, that standard model neutrinos are produced as left
polarized particles. If they gravitationally interact with BH, some
incoming left neutrinos become right polarized after scattering. A
neutrino detector can observe only left neutrinos. Hence, the observed
flux of neutrinos is $F_{\nu}=P_{\mathrm{LL}}F_{0}$, where $F_{0}$
is the flux of scalar particles. The value of $F_{0}$ is proportional
to the differential cross section, $F_{0}\sim\mathrm{d}\sigma/\mathrm{d}\varOmega$,
which is studied in Appendix~\ref{sec:PARTM}.

We assume that the neutrino beam scatters off a SMBH surrounded by
an accretion disk. For example, we can suppose that such a SMBH is
in the center of a Seyfert galaxy. We take that the plasma density
in the disk scales as $n_{e}\propto r^{-\beta}$. The value of $\beta$
is very model dependent. For example, $\beta\approx0.5$ in an advection
dominated accretion disk studied in Ref.~\cite{Igu00}. If we take
that the mass of SMBH in question is $M\sim10^{8}M_{\odot}$, the
plasma density in the vicinity of SMBH can be up to $n_{e}\sim10^{18}\,\text{cm}^{-3}$~\cite{Jia19}.
Thus, the dimensionless effective potential $V(r)=G_{F}n_{e}(r)r_{g}/\sqrt{2}$,
reads $V(x)=V_{\mathrm{max}}x^{-\beta}$, where $x=r/r_{g}$.

One can consider various neutrino trajectories with respect to an
accretion disk. However, to highlight the effect of the neutrino interaction
with matter we study two extreme cases: the neutrino motion in the
plane perpendicular to an accretion disk, marked by the symbol $\perp$,
and the neutrino propagation in the plane of an accretion disk, labeled
by the symbol $\parallel$. The effect of the neutrino matter interaction
is maximal in the latter situation since we assume that a disk is
slim.

\begin{figure}
  \centering
  \subfigure[]
  {\label{1a}
  \includegraphics[scale=.35]{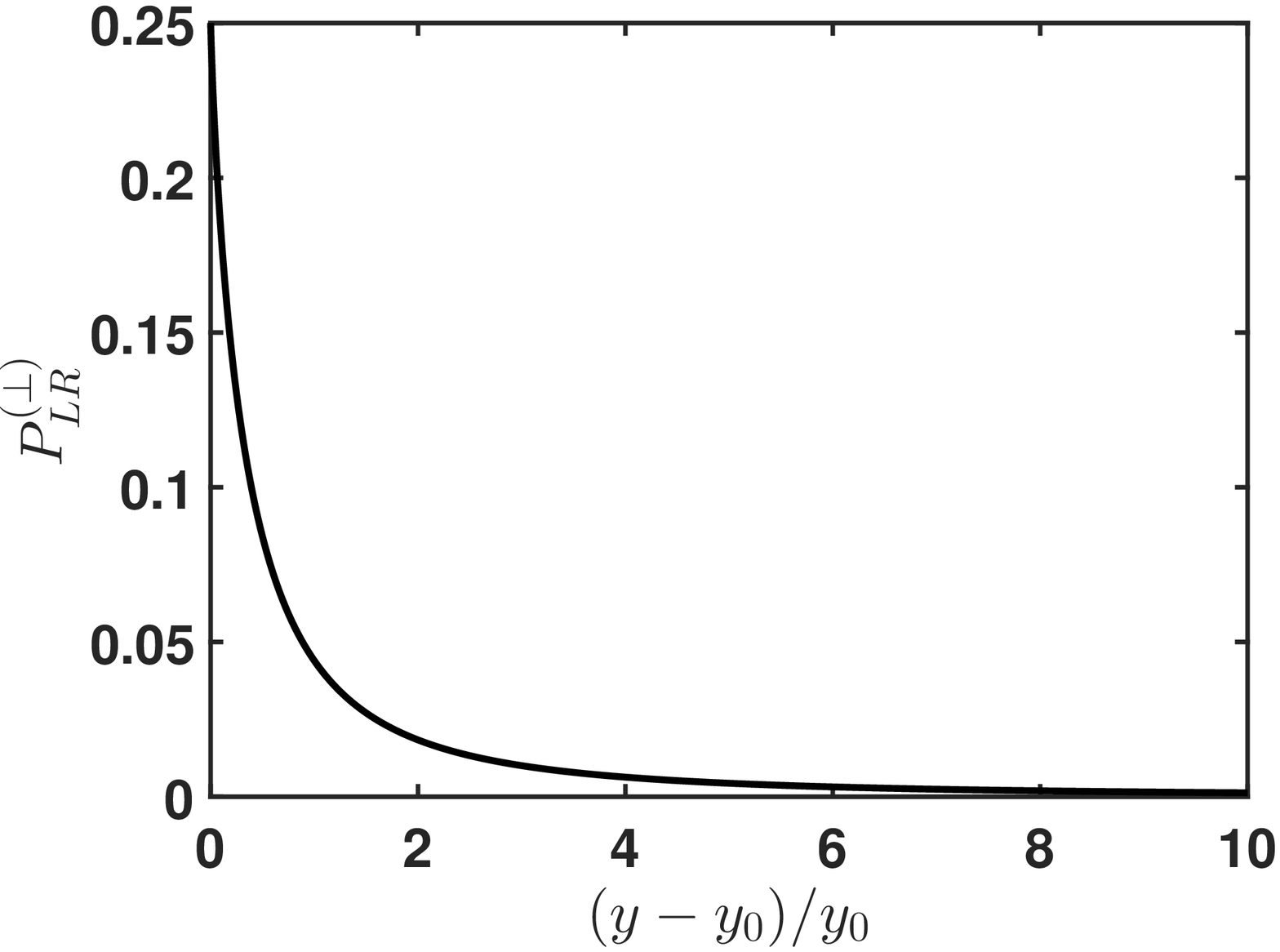}}
  \hskip-.6cm
  \subfigure[]
  {\label{1b}
  \includegraphics[scale=.35]{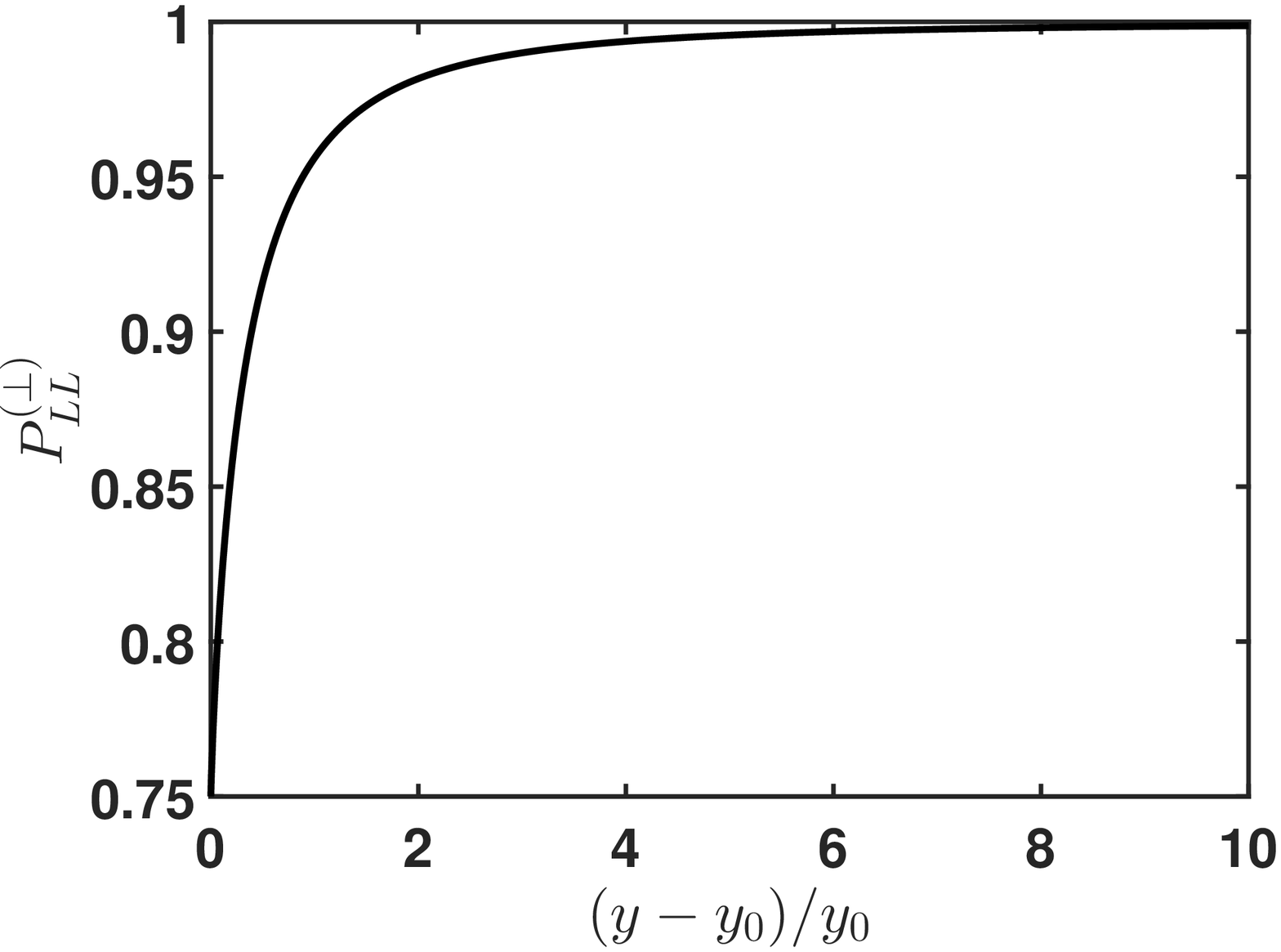}}
  \\
  \subfigure[]
  {\label{1c}
  \includegraphics[scale=.35]{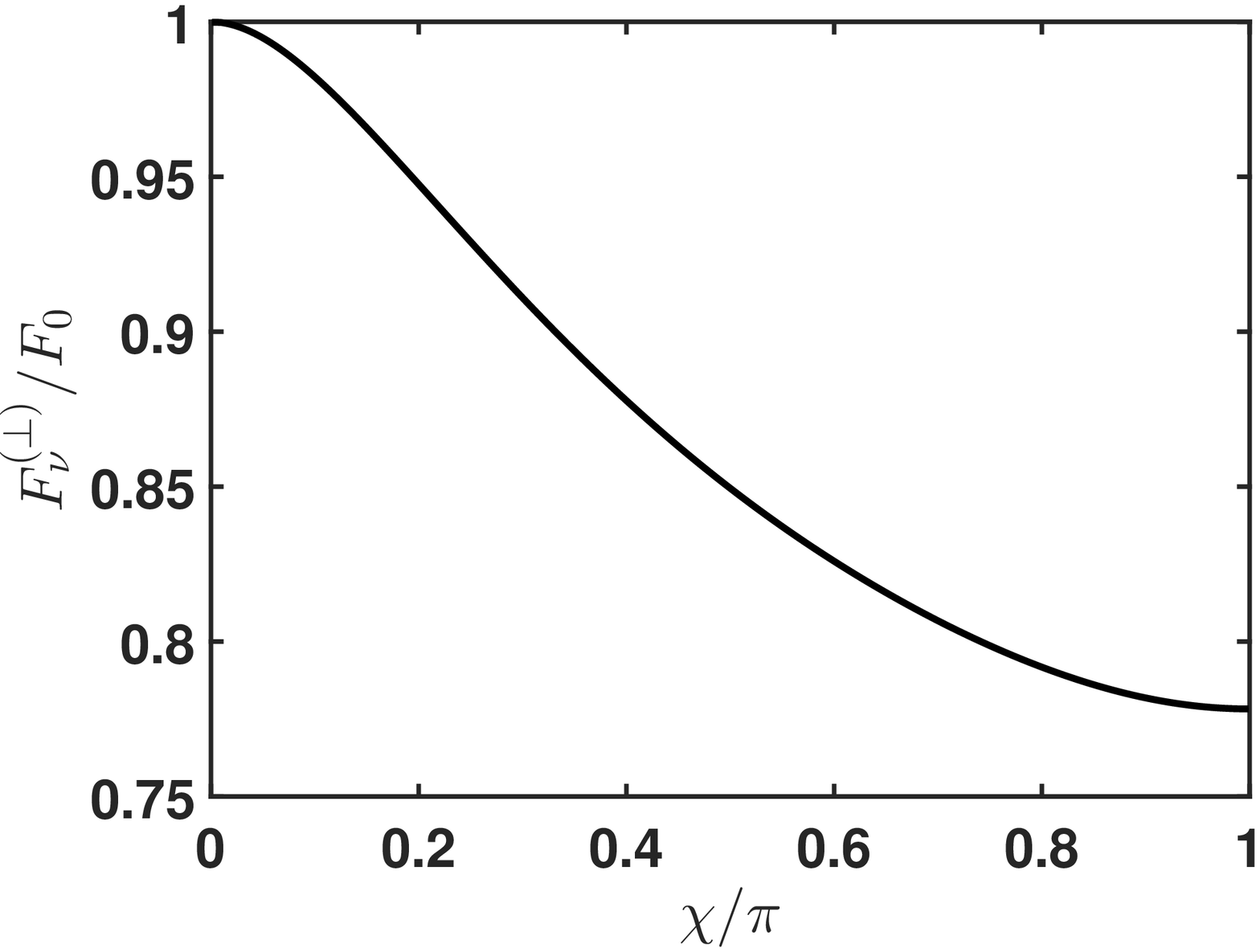}}
  \protect
  \caption{(a)~The transition probability of spin oscillations $P_{\mathrm{LR}}^{(\perp)}$,
  when a neutrino moves perpendicularly to an accretion disk, versus
  the dimensionless impact parameter $y$. (b)~The survival probability
  of spin oscillations $P_{\mathrm{LL}}^{(\perp)}$, when a neutrino
  moves perpendicularly to an accretion disk, as a function the dimensionless
  impact parameter $y$. (c)~The ratio of the measured fluxes of neutrinos
  and scalar particles, when they move perpendicularly to an accretion
  disk, versus the scattering angle $\chi$ normalized by $\pi$.\label{fig:perpscat}}
\end{figure}

First we discuss the case $\perp$, when only the gravity contributes
the neutrino scattering off BH. The transition and survival probabilities,
as the functions of the dimensionless impact parameter $y=b/r_{g}$,
are shown in Figs.~\ref{fig:perpscat}(a) and~\ref{fig:perpscat}(b).
Despite we show the probabilities for $y_{0}<y<11y_{0}$ (see also
Figs.~\ref{fig:paralscatLR} and~\ref{fig:paralscatLL} below),
we take that $y<30y_{0}$ in our simulations. One can see in Figs.~\ref{fig:perpscat}(a)
and~\ref{fig:perpscat}(b) that $P_{\mathrm{LR}}^{(\perp)}\to0.25$
($P_{\mathrm{LL}}^{(\perp)}\to0.75$) at $y\to y_{0}$ and $P_{\mathrm{LR}}^{(\perp)}\to0$
($P_{\mathrm{LL}}^{(\perp)}\to1$) at $y\gg y_{0}$, which is in agreement
with the results in Sec.~\ref{sec:GRAV}.

The probabilities as functions of the impact parameter, shown in Figs.~\ref{1a}
and~\ref{1b}, are not the measurable quantities. In
Fig.~\ref{1c}, we show the measured flux of neutrinos,
moving perpendicularly to an accretion disk, $F_{\nu}^{(\perp)}$,
normalized by the flux of scalar particles, versus the scattering
angle $\chi$. These fluxes are proportional to the differential cross
section, $F\sim\mathrm{d}\sigma/\mathrm{d}\varOmega$, where $\mathrm{d}\varOmega=2\pi\sin\chi\mathrm{d}\chi$.
The flux of scalar particles (or the cross section), $F_{0}$, is
given in Appendix (see also Ref.~\cite{DolDorLas06}).

One can see in Fig.~\ref{1c} that spin effects in the
neutrino gravitational scattering off BH significantly reduce the
observed flux of neutrinos compared to the case of scalar particles.
The influence of spin oscillations is maximal for neutrinos scattered
backwards. The reduction of the flux can be more than 20\% in this
situation.

Now we turn to the discussion of the neutrino interaction with both
the gravity and an accretion disk, i.e. we discuss the case $\parallel$.
The transition and survival probabilities are shown on Figs.~\ref{fig:paralscatLR}
and~\ref{fig:paralscatLL} for various $V_{\mathrm{max}}$, or maximal
plasma density $n_{e}$, and $\beta$.

One can see that the best coincidence between $\perp$ and $\parallel$
cases is implemented when $V_{\mathrm{max}}=0.1$ and $\beta=0.5$;
cf. Figs.~\ref{1a} and~\ref{2a},
as well as Figs.~\ref{1b} and~\ref{3a}.
Indeed, this situation corresponds to a low density accretion disk
with $n_{e}=10^{18}\,\text{cm}^{-3}$, which has relatively rapid
density decrease (great $\beta=0.5$), i.e. the influence of the neutrino
matter interaction is minimal. The opposite case is presented in Figs.
~\ref{2d} and~\ref{3d}, where
the influence of matter on neutrino spin oscillations is maximal since
matter density is higher, $n_{e}=2\times10^{18}\,\text{cm}^{-3}$.
Moreover, the density profile is less steep (small $\beta=0.2$),
i.e. a neutrino stays longer inside such a disk.

\begin{figure}
  \centering
  \subfigure[]
  {\label{2a}
  \includegraphics[scale=.35]{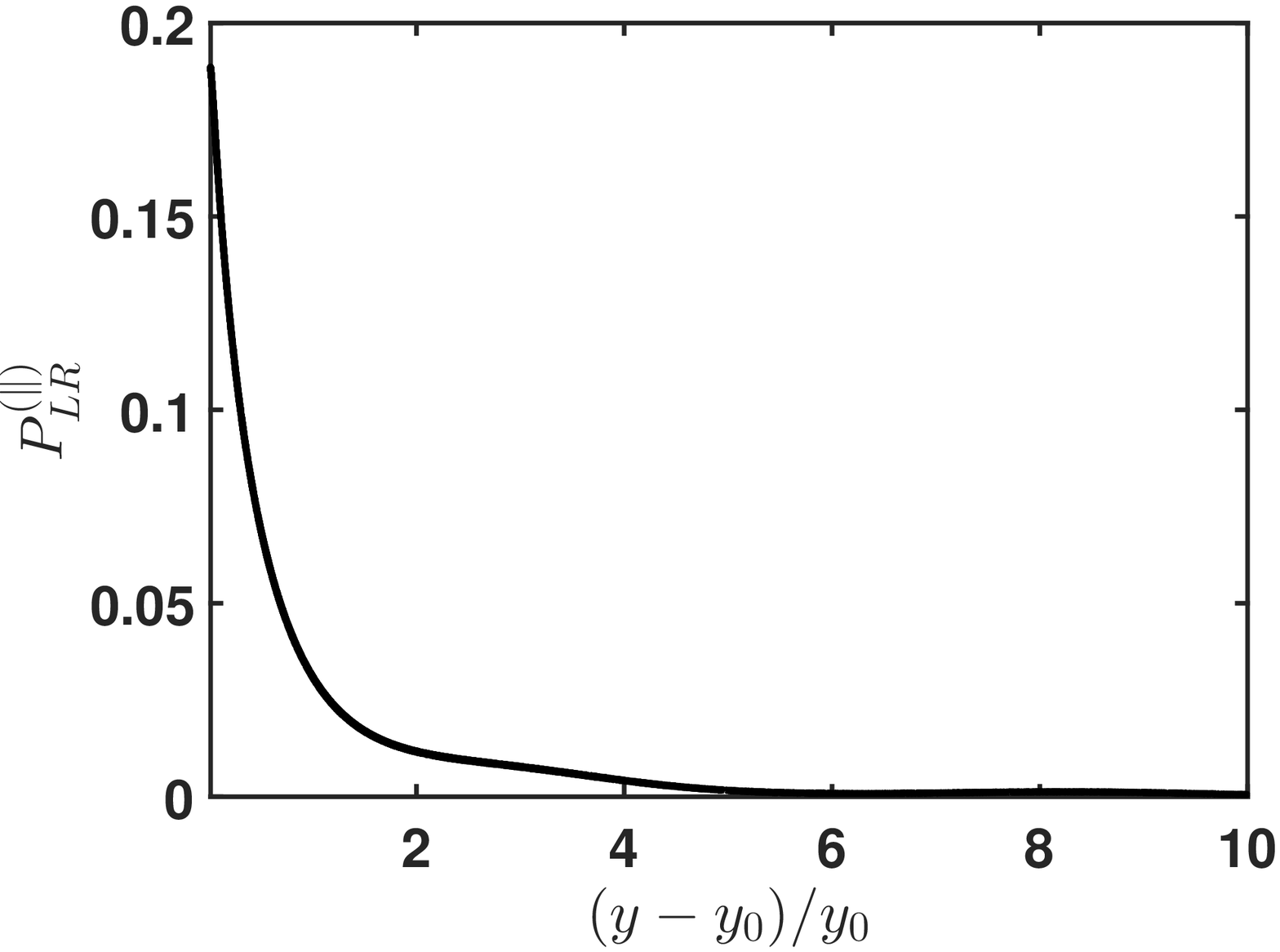}}
  \hskip-.6cm
  \subfigure[]
  {\label{2b}
  \includegraphics[scale=.35]{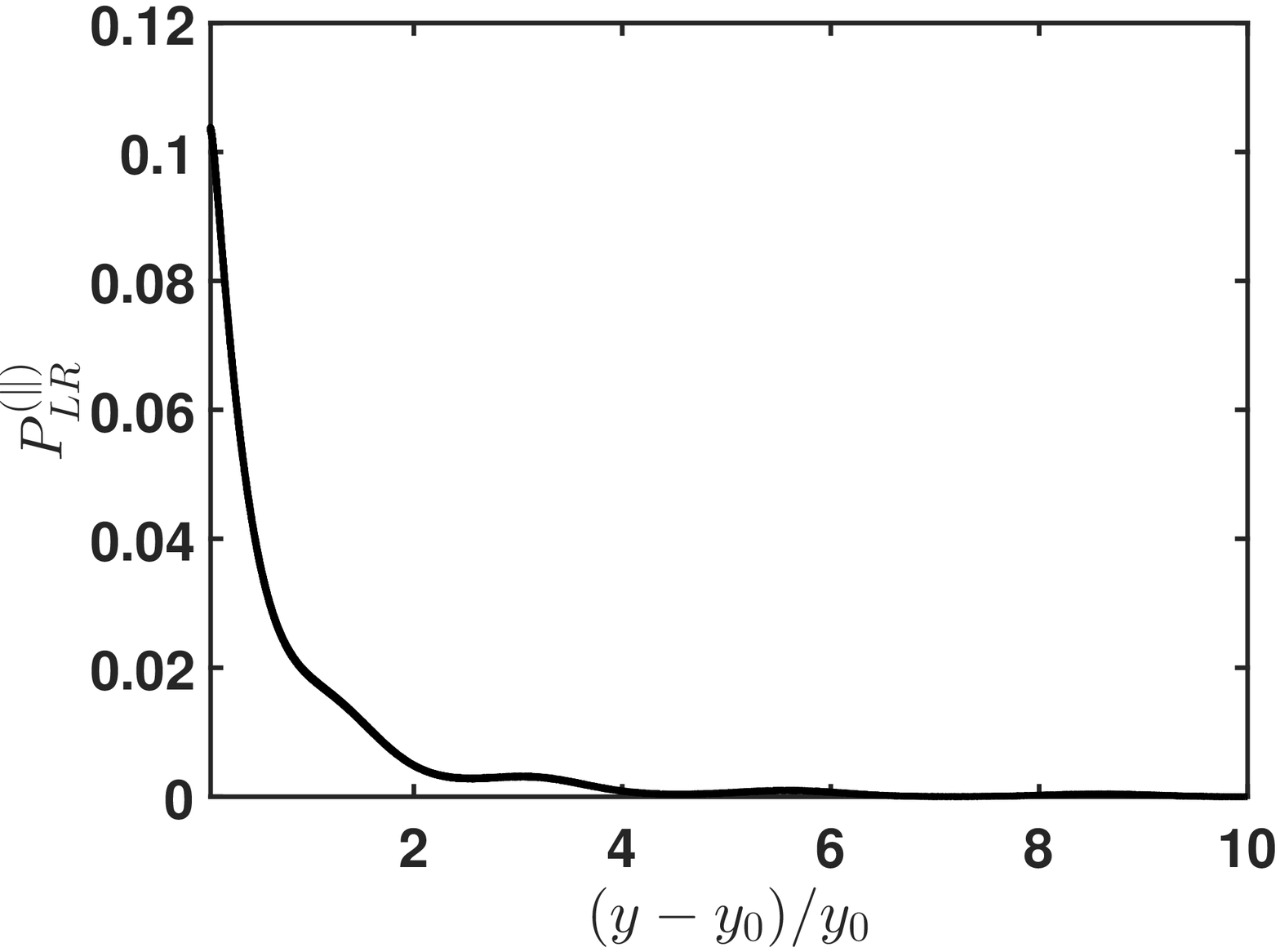}}
  \\
  \subfigure[]
  {\label{2c}
  \includegraphics[scale=.35]{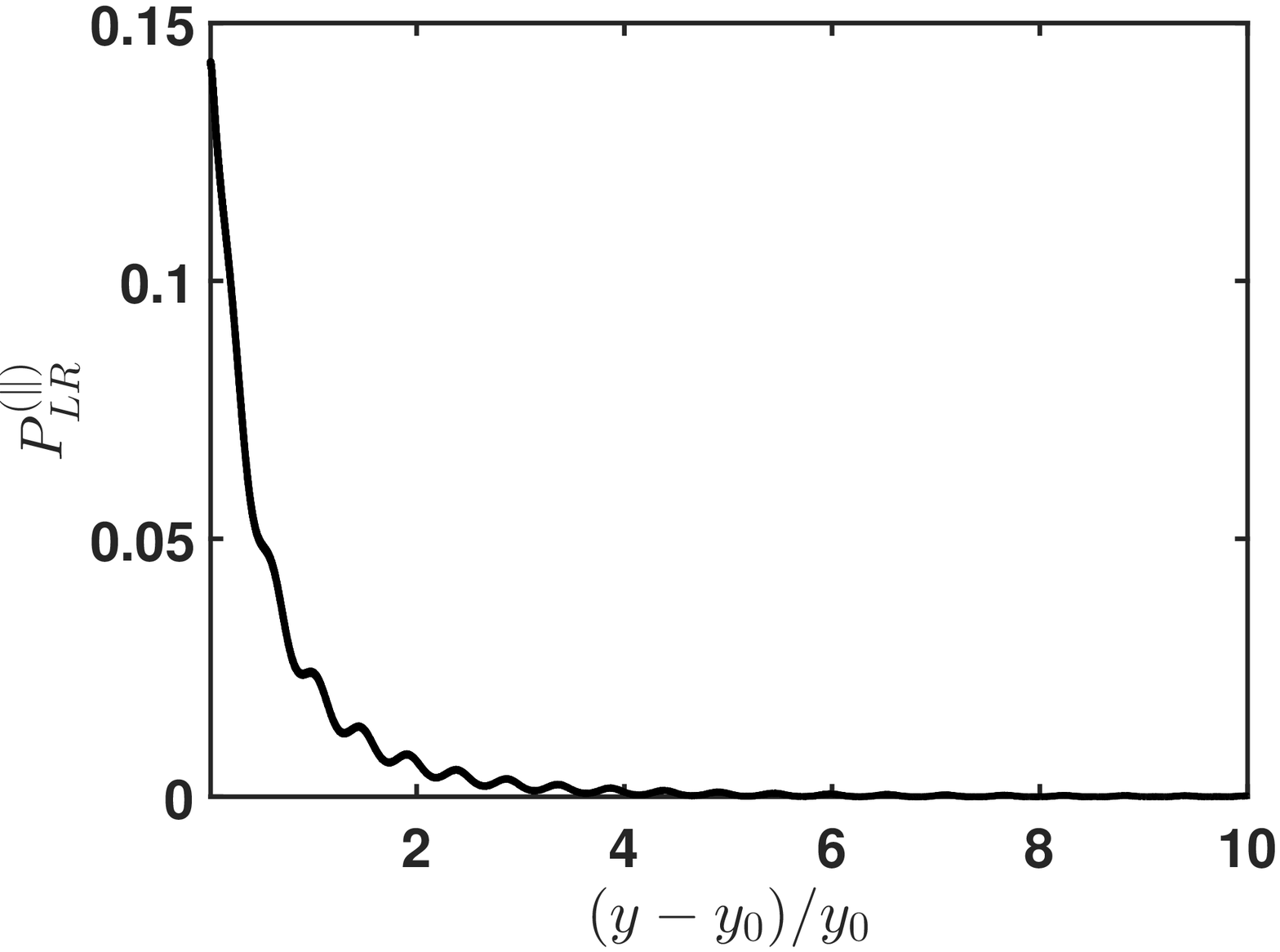}}
  \hskip-.6cm
  \subfigure[]
  {\label{2d}
  \includegraphics[scale=.35]{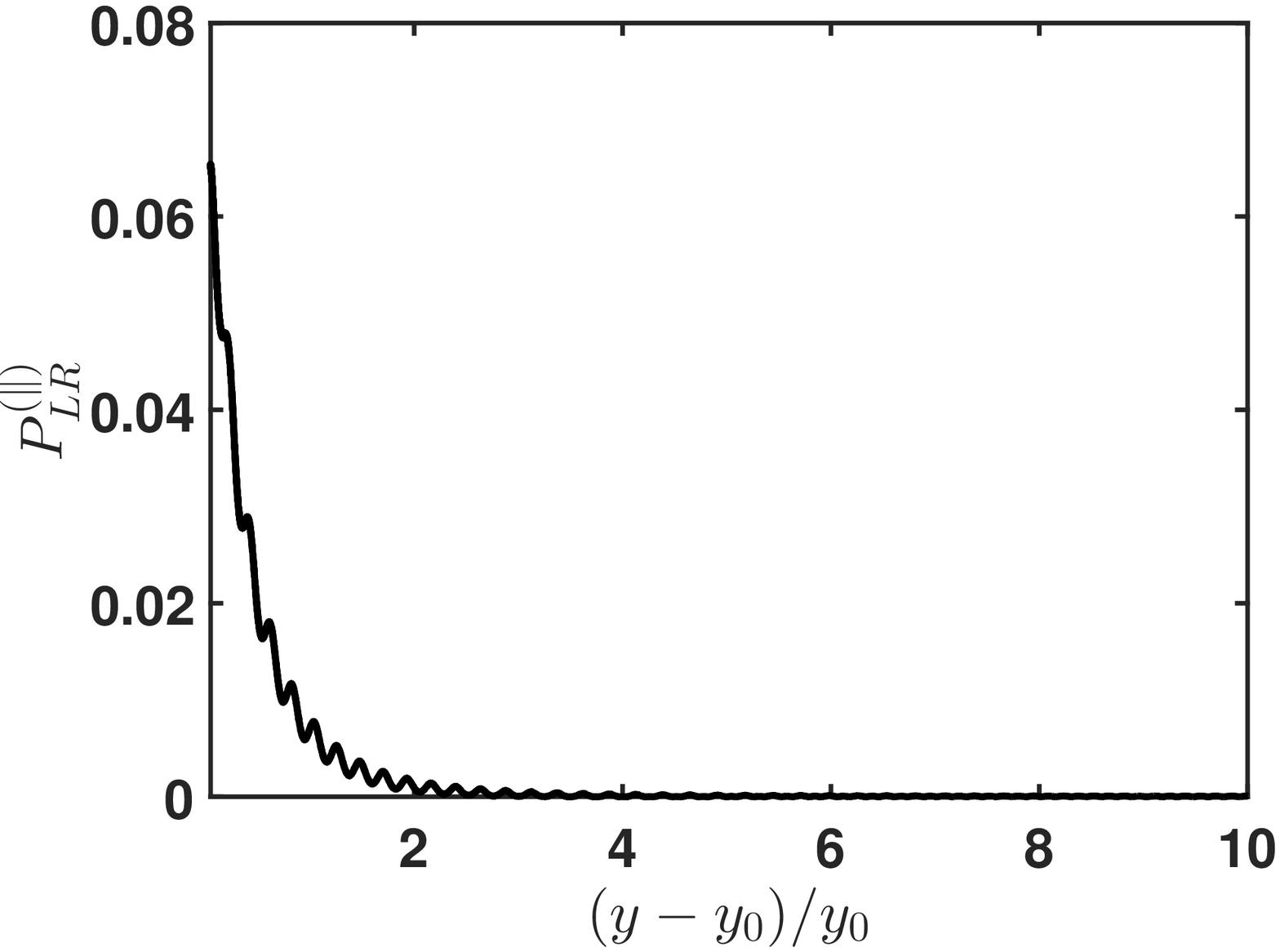}}
  \protect
  \caption{The transition probability of spin oscillations $P_{\mathrm{LR}}^{(\parallel)}$,
  when a neutrino interacts with an accretion disk, versus the dimensionless
  impact parameter $y$. (a)~$V_{\mathrm{max}}=0.1$ ($n_{e}=10^{18}\,\text{cm}^{-3}$)
  and $\beta=0.5$; (b)~$V_{\mathrm{max}}=0.2$ ($n_{e}=2\times10^{18}\,\text{cm}^{-3}$)
  and $\beta=0.5$; (c)~$V_{\mathrm{max}}=0.1$ and $\beta=0.2$;
  (d)~$V_{\mathrm{max}}=0.2$ and $\beta=0.2$.\label{fig:paralscatLR}}
\end{figure}

\begin{figure}
  \centering
  \subfigure[]
  {\label{3a}
  \includegraphics[scale=.35]{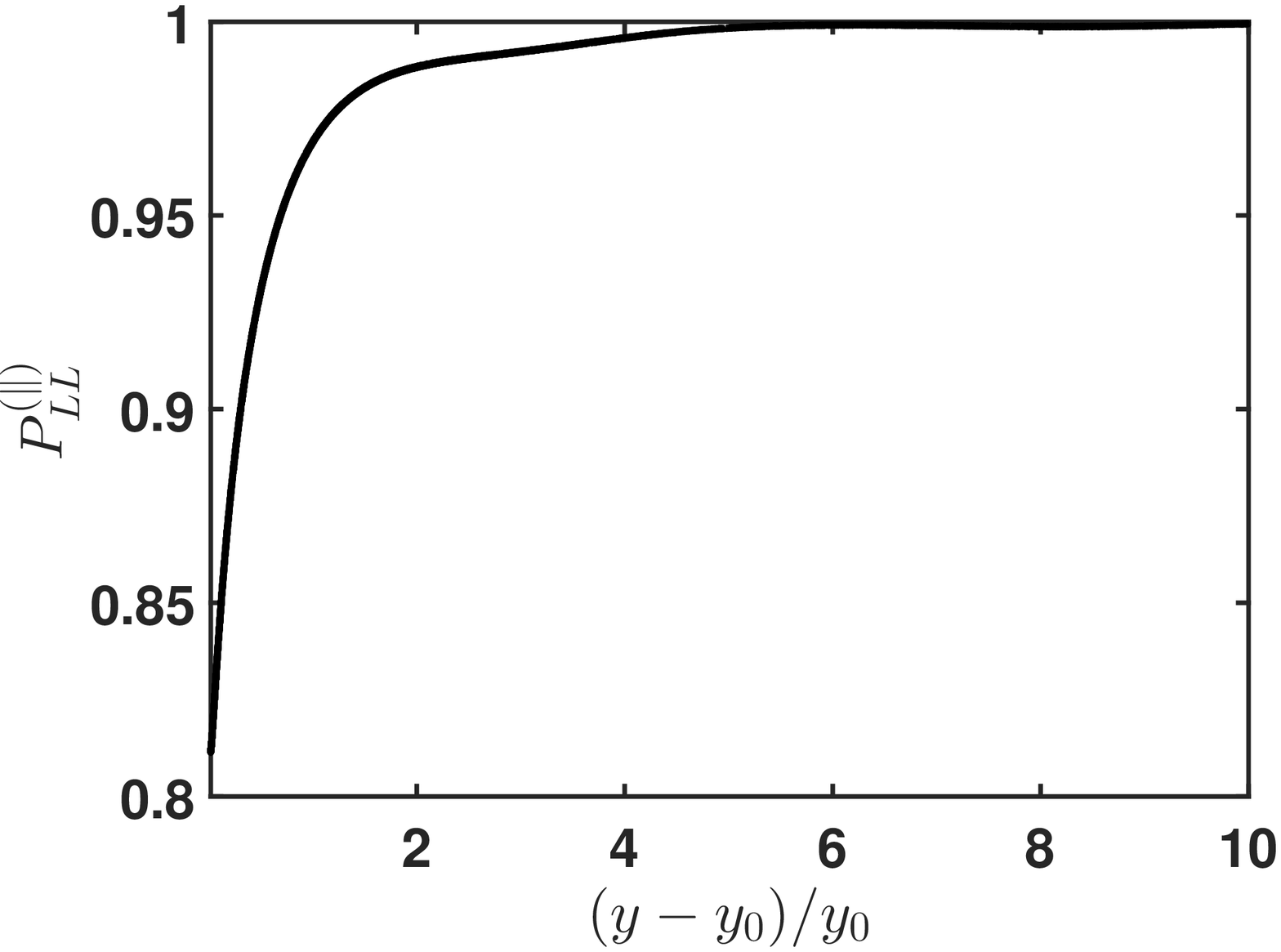}}
  \hskip-.6cm
  \subfigure[]
  {\label{3b}
  \includegraphics[scale=.35]{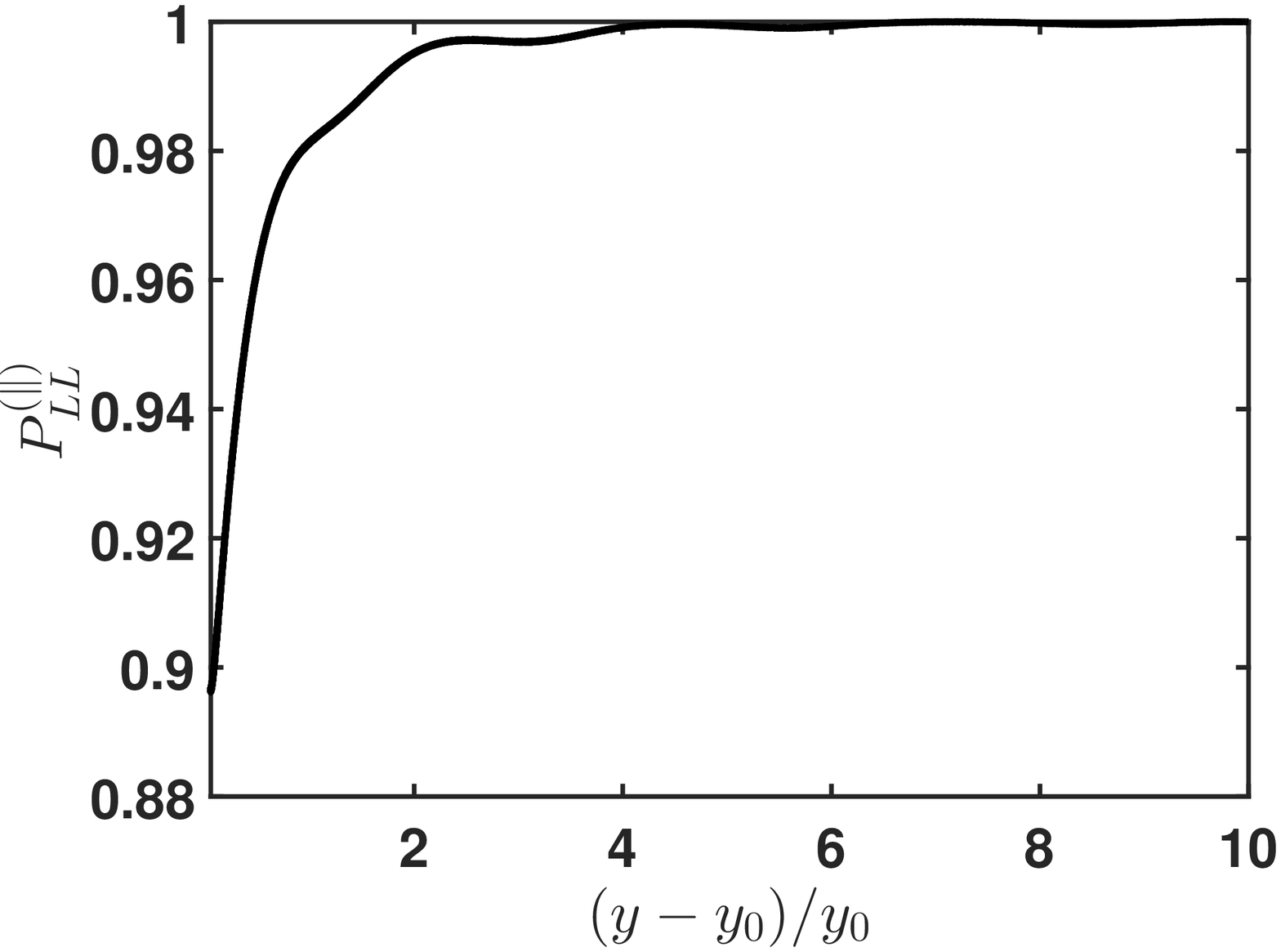}}
  \\
  \subfigure[]
  {\label{3c}
  \includegraphics[scale=.35]{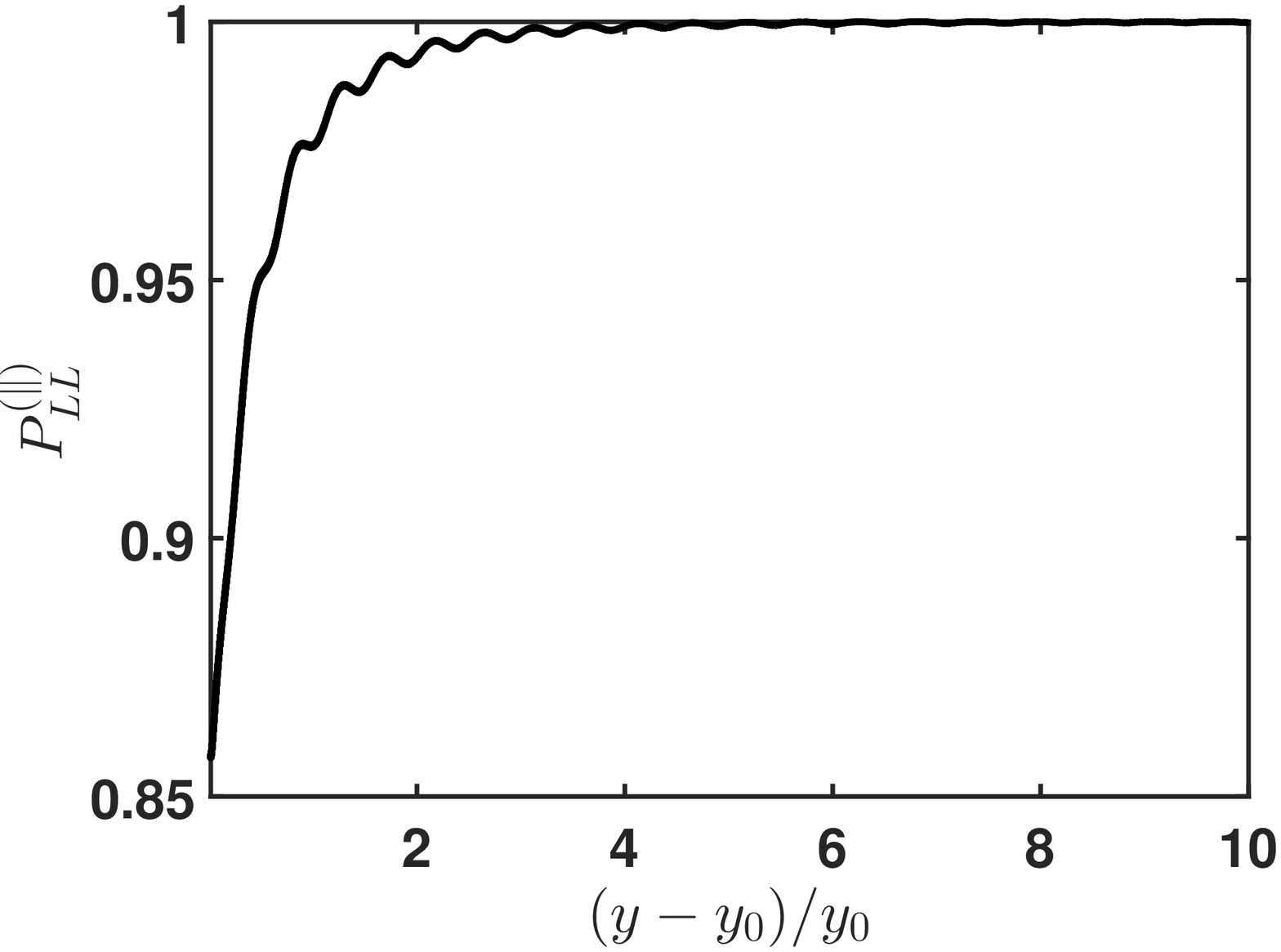}}
  \hskip-.6cm
  \subfigure[]
  {\label{3d}
  \includegraphics[scale=.35]{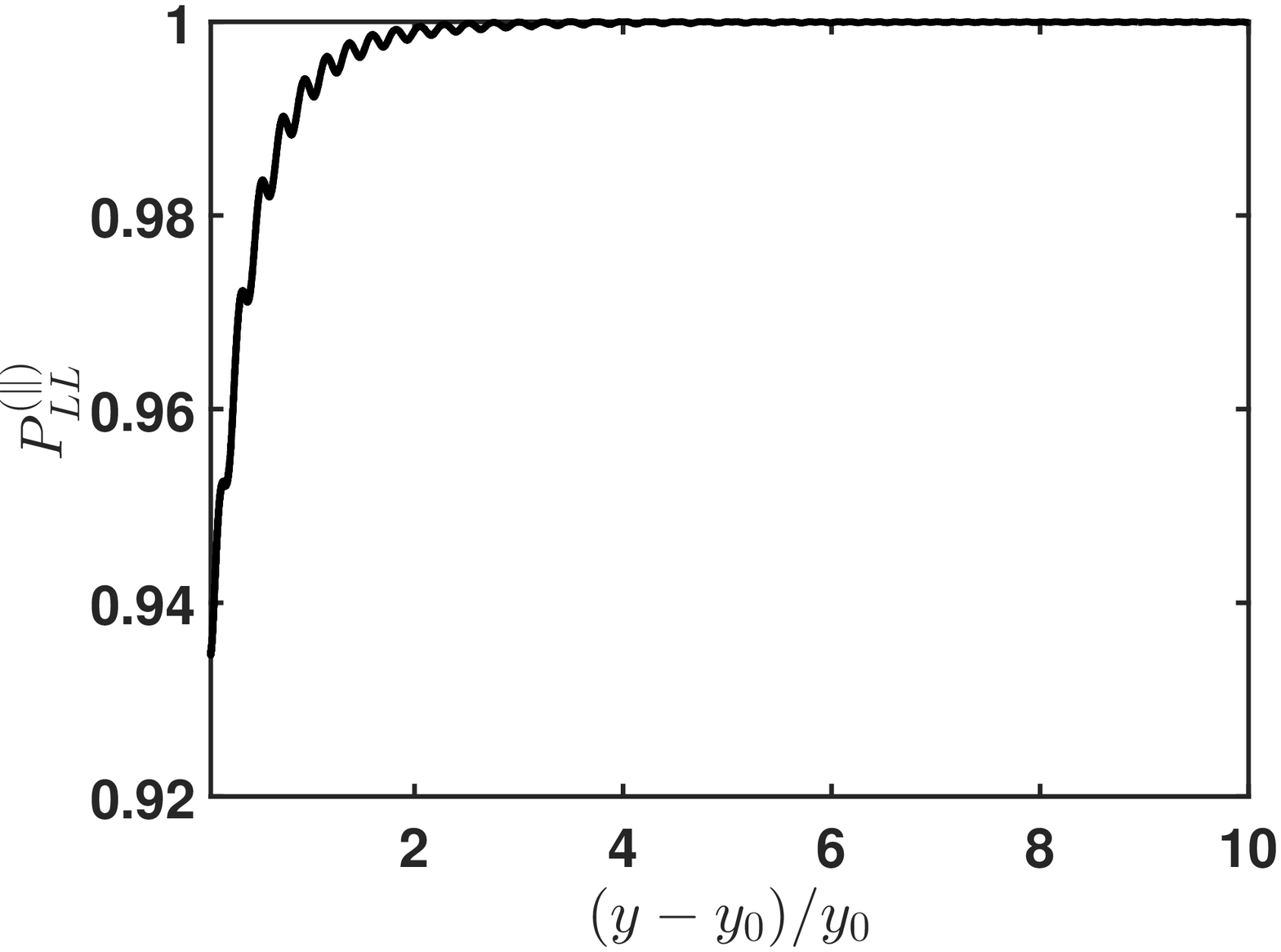}}
  \protect
  \caption{The survival probability of spin oscillations $P_{\mathrm{LL}}^{(\parallel)}$,
  when a neutrino interacts with an accretion disk, versus the dimensionless
  impact parameter $y$ for different $V_{\mathrm{max}}$ and $\beta$.
  (a)~$V_{\mathrm{max}}=0.1$ ($n_{e}=10^{18}\,\text{cm}^{-3}$) and
  $\beta=0.5$; (b)~$V_{\mathrm{max}}=0.2$ ($n_{e}=2\times10^{18}\,\text{cm}^{-3}$)
  and $\beta=0.5$; (c)~$V_{\mathrm{max}}=0.1$ and $\beta=0.2$;
  (d)~$V_{\mathrm{max}}=0.2$ and $\beta=0.2$.\label{fig:paralscatLL}}
\end{figure}

We also mention that the analysis of neutrino spin oscillations in
accretion disks with a constant density, studied in Ref.~\cite{Jia19},
is problematic because of difficulties in the numerical solution of
Eqs.~(\ref{eq:Schr}) and~(\ref{eq:Omegar}) in the limit $\beta\to0$.

\begin{figure}
  \centering
  \subfigure[]
  {\label{4a}
  \includegraphics[scale=.35]{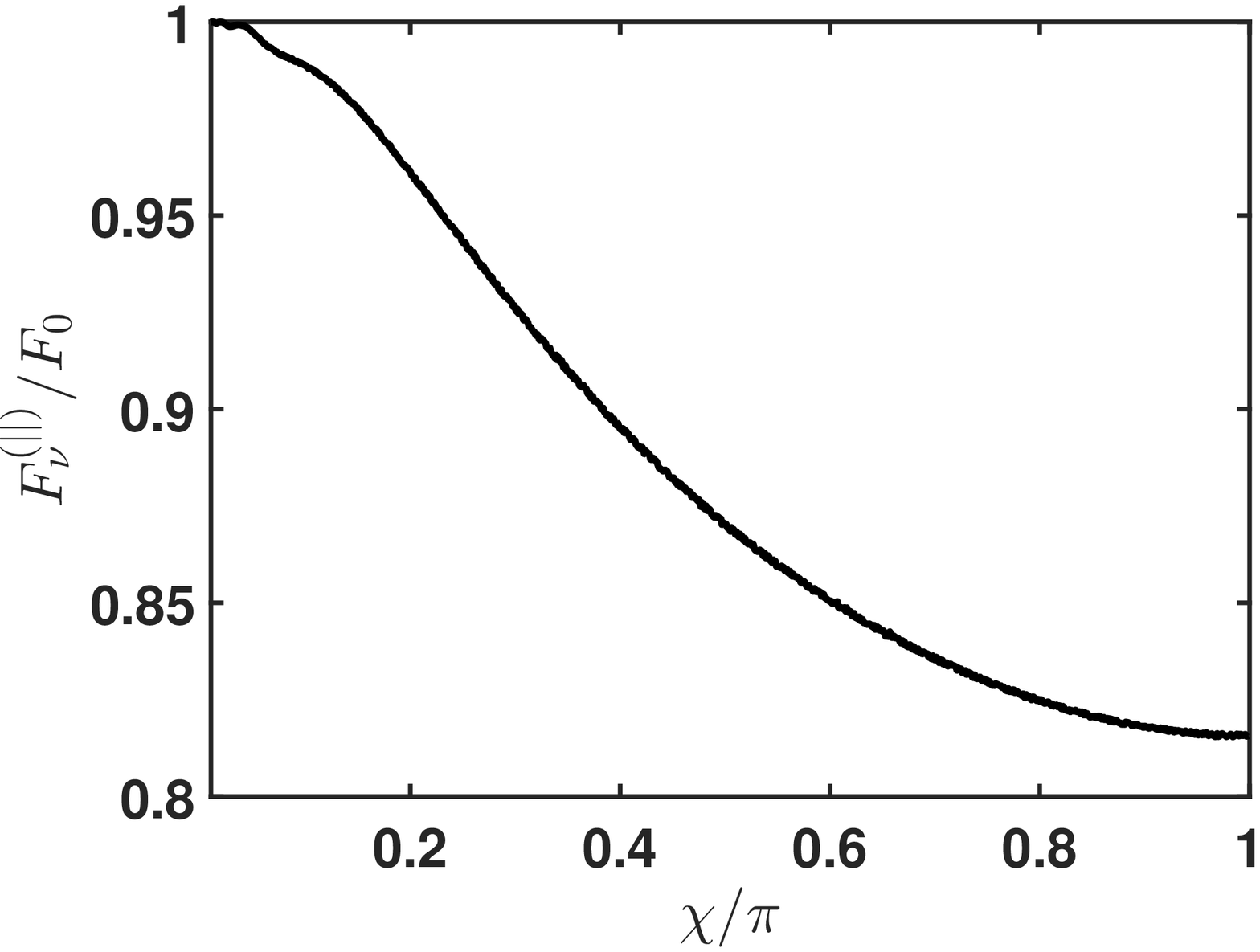}}
  \hskip-.6cm
  \subfigure[]
  {\label{4b}
  \includegraphics[scale=.35]{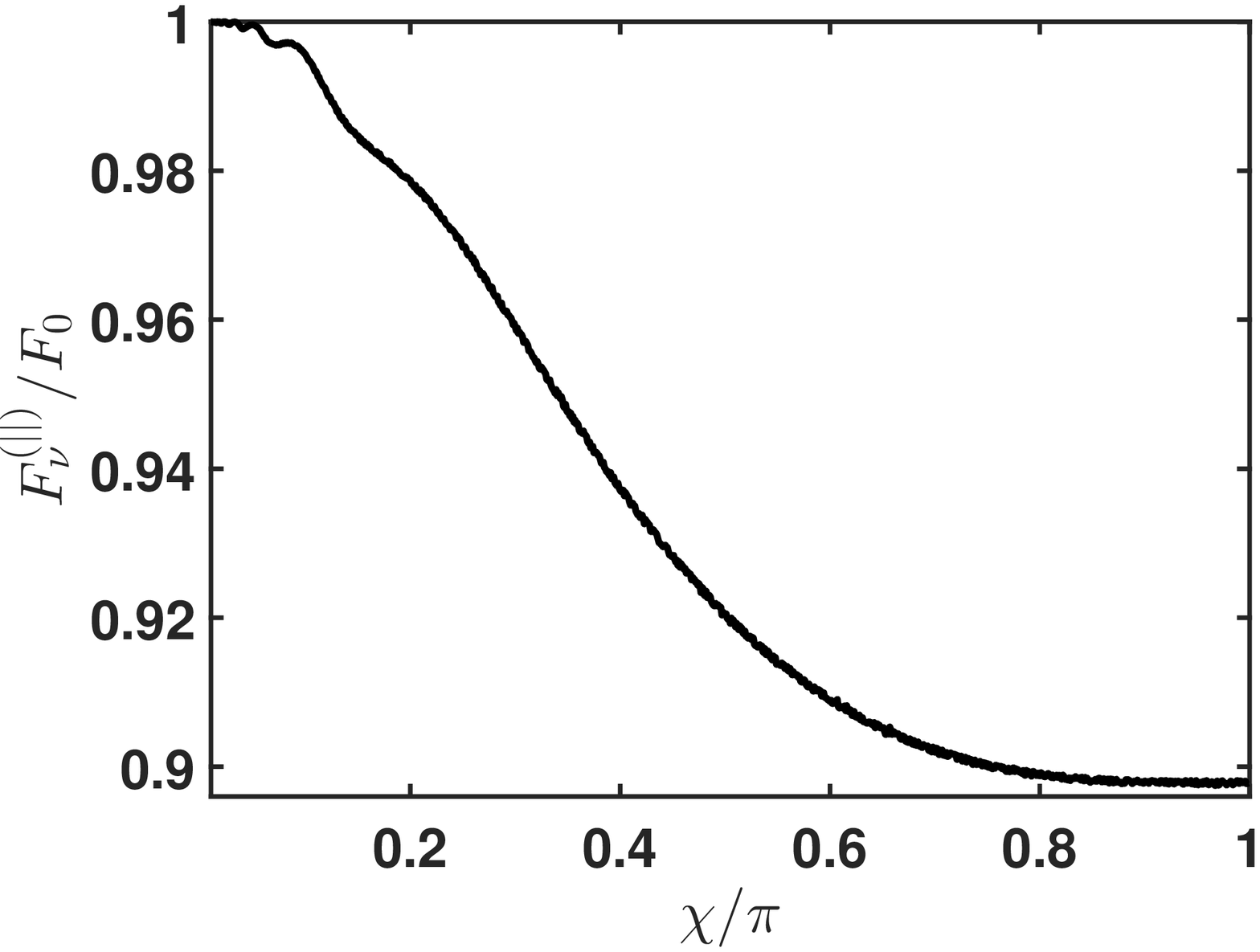}}
  \\
  \subfigure[]
  {\label{4c}
  \includegraphics[scale=.35]{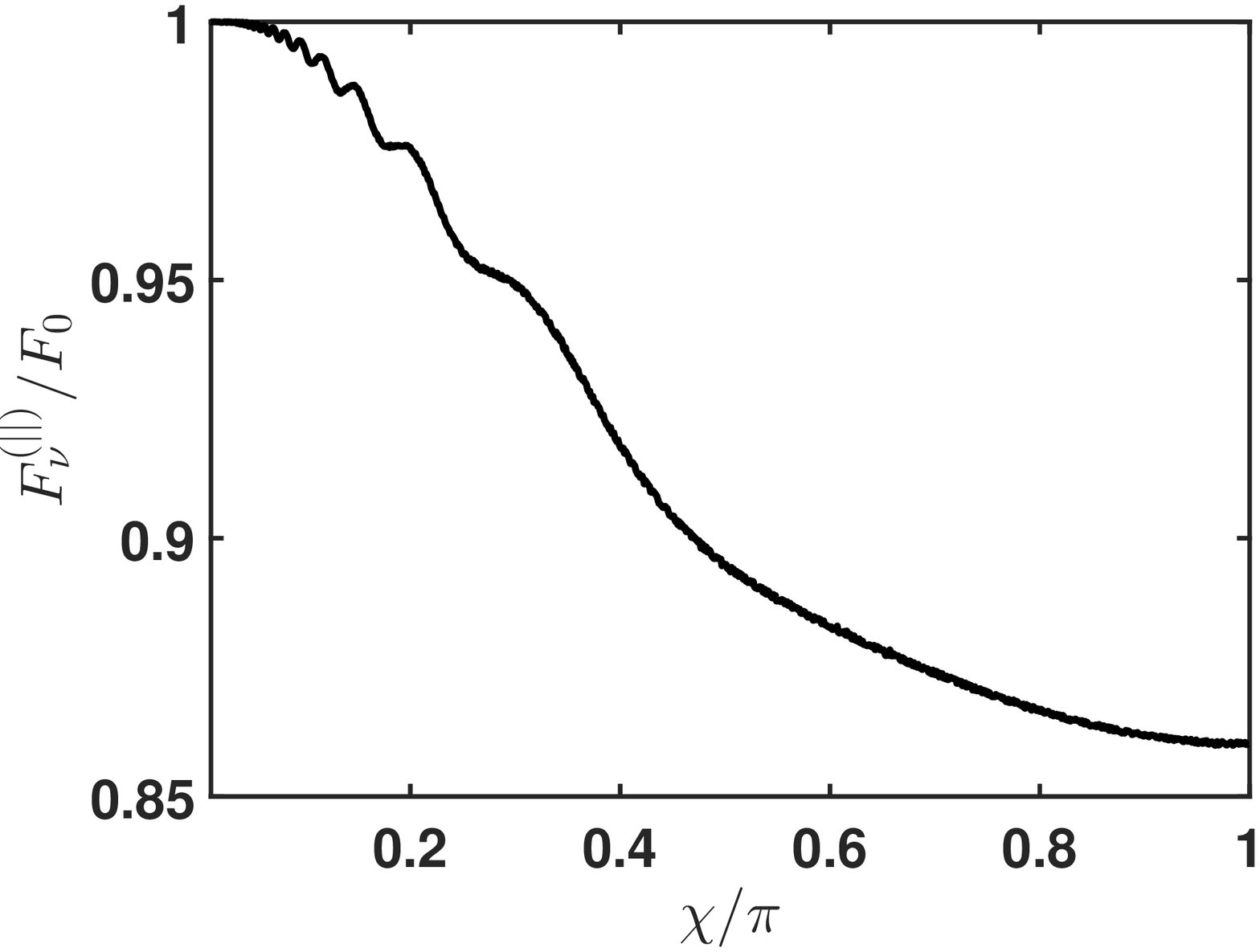}}
  \hskip-.6cm
  \subfigure[]
  {\label{4d}
  \includegraphics[scale=.35]{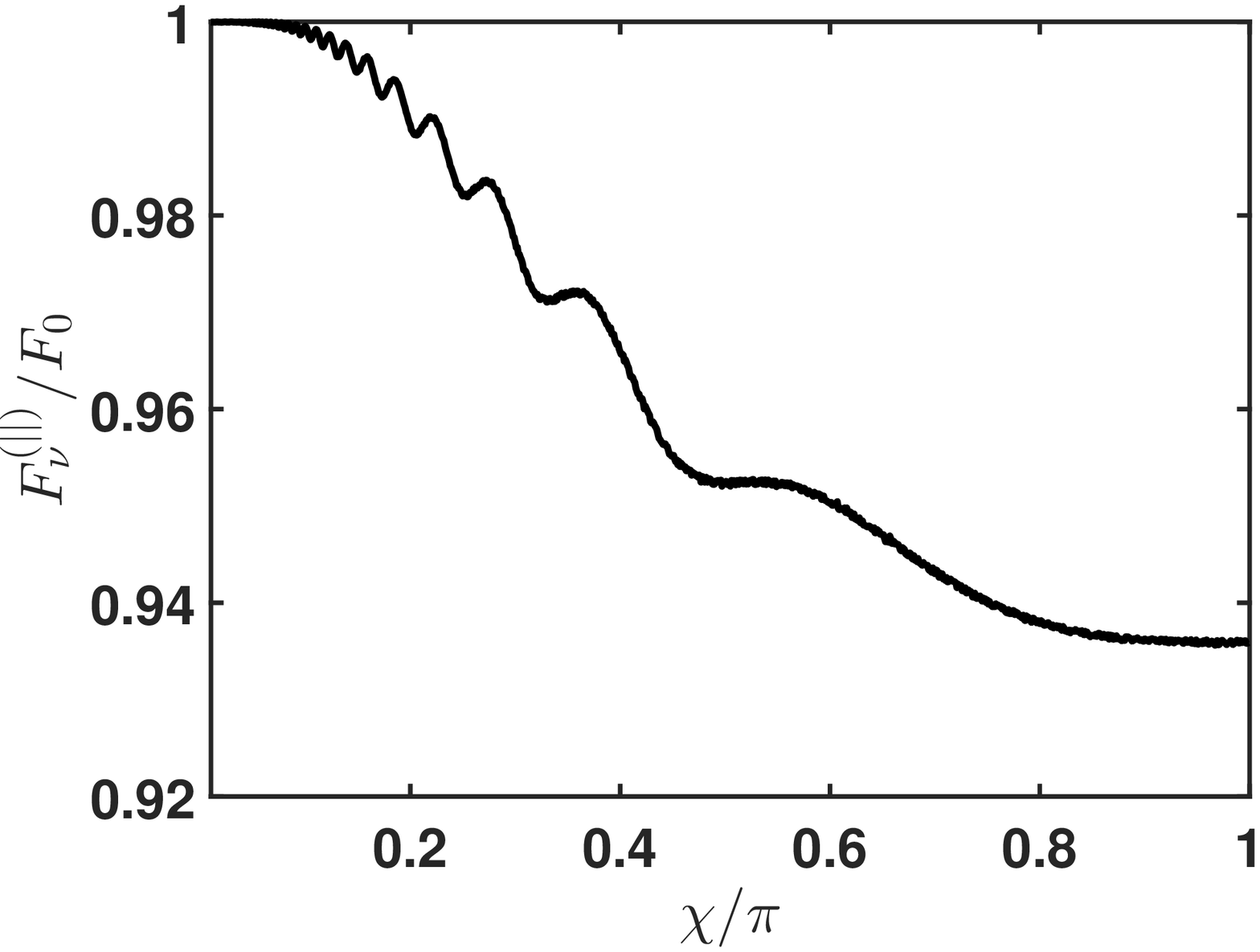}}
  \protect
  \caption{The fluxes of neutrinos, normalized by the flux of scalar particles,
  for particles interacting with background matter of an accretion disk.
  Here we represent the dependence of $F_{\nu}^{(\parallel)}$ on $\chi$
  for different $V_{\mathrm{max}}$ and $\beta$. (a)~$V_{\mathrm{max}}=0.1$
  ($n_{e}=10^{18}\,\text{cm}^{-3}$) and $\beta=0.5$; (b)~$V_{\mathrm{max}}=0.2$
  ($n_{e}=2\times10^{18}\,\text{cm}^{-3}$) and $\beta=0.5$; (c)~$V_{\mathrm{max}}=0.1$
  and $\beta=0.2$; (d)~$V_{\mathrm{max}}=0.2$ and $\beta=0.2$.\label{fig:Fparal}}
\end{figure}

We have already mentioned above that the probabilities of spin oscillations
versus the impact parameter cannot be measured in an experiment. In
Fig.~\ref{fig:Fparal}, we show the fluxes of neutrinos $F_{\nu}^{(\parallel)}$
scattered off BH and interacting with an accretion disk. These fluxes
are normalized to the flux of scalar particles. As mentioned above,
the best coincidence between $\perp$ and $\parallel$ cases is implemented
for $V_{\mathrm{max}}=0.1$ and $\beta=0.5$; cf. Figs.~\ref{4a}
and~\ref{2c}.

\begin{figure}
  \centering
  \subfigure[]
  {\label{5a}
  \includegraphics[scale=.35]{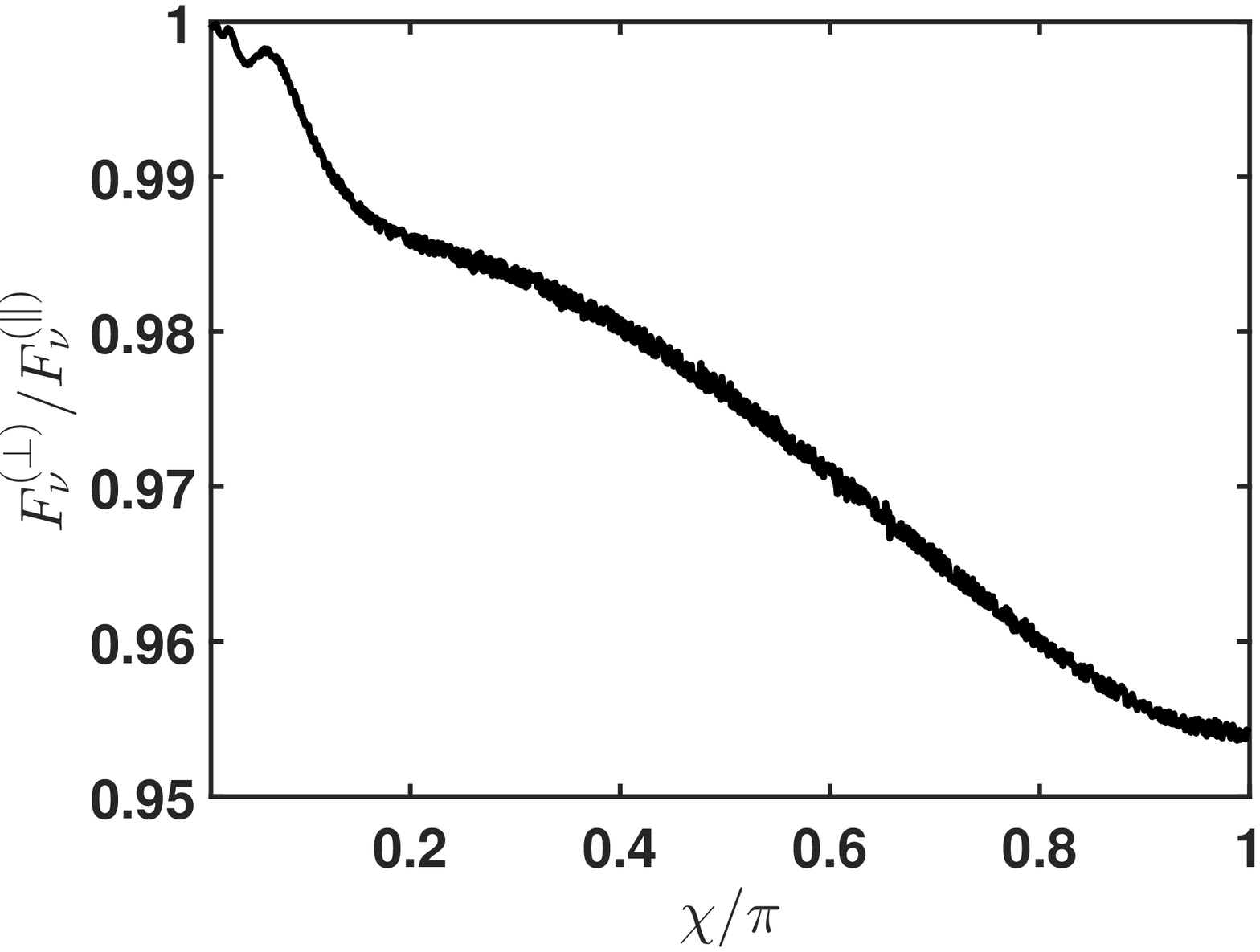}}
  \hskip-.6cm
  \subfigure[]
  {\label{5b}
  \includegraphics[scale=.35]{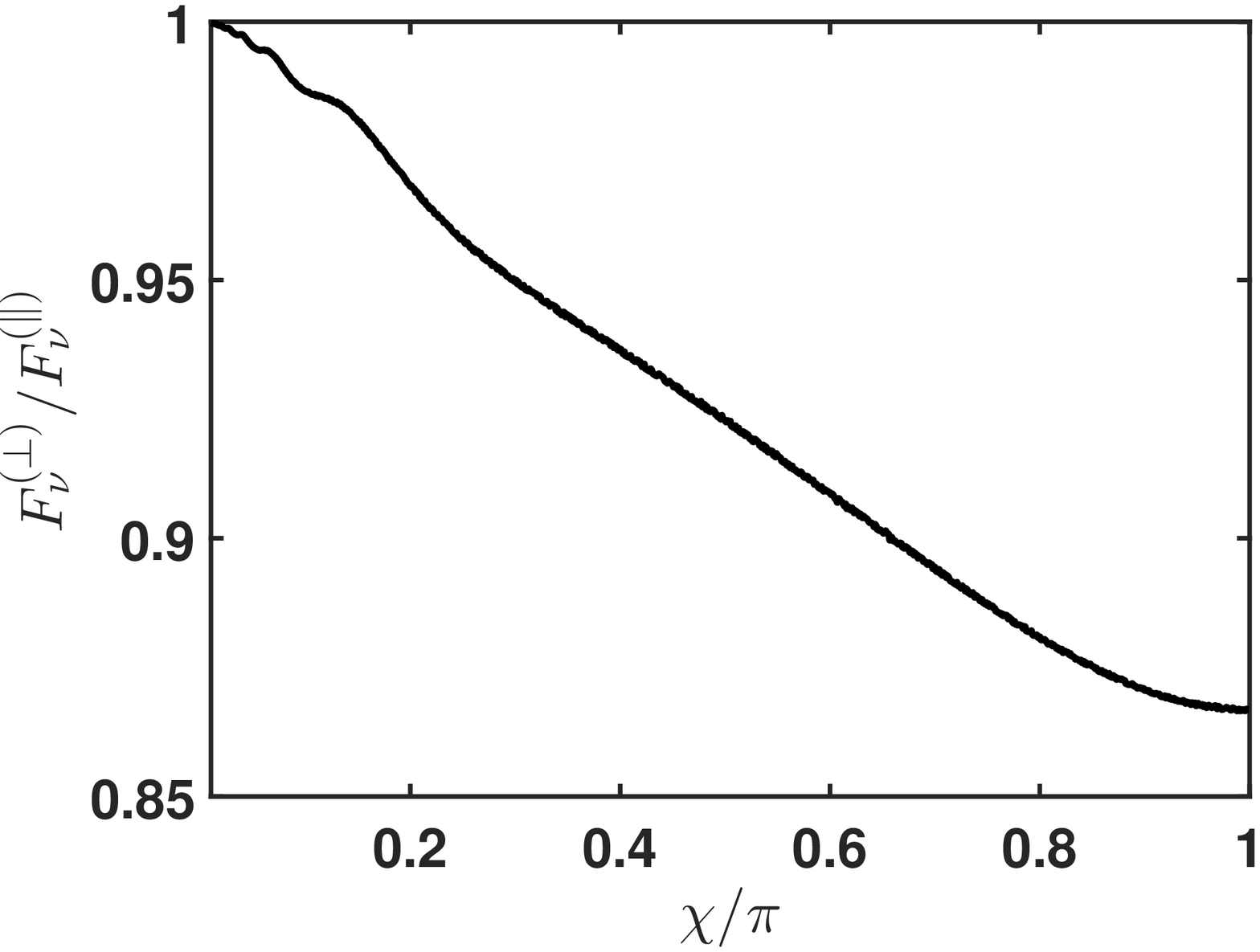}}
  \\
  \subfigure[]
  {\label{5c}
  \includegraphics[scale=.35]{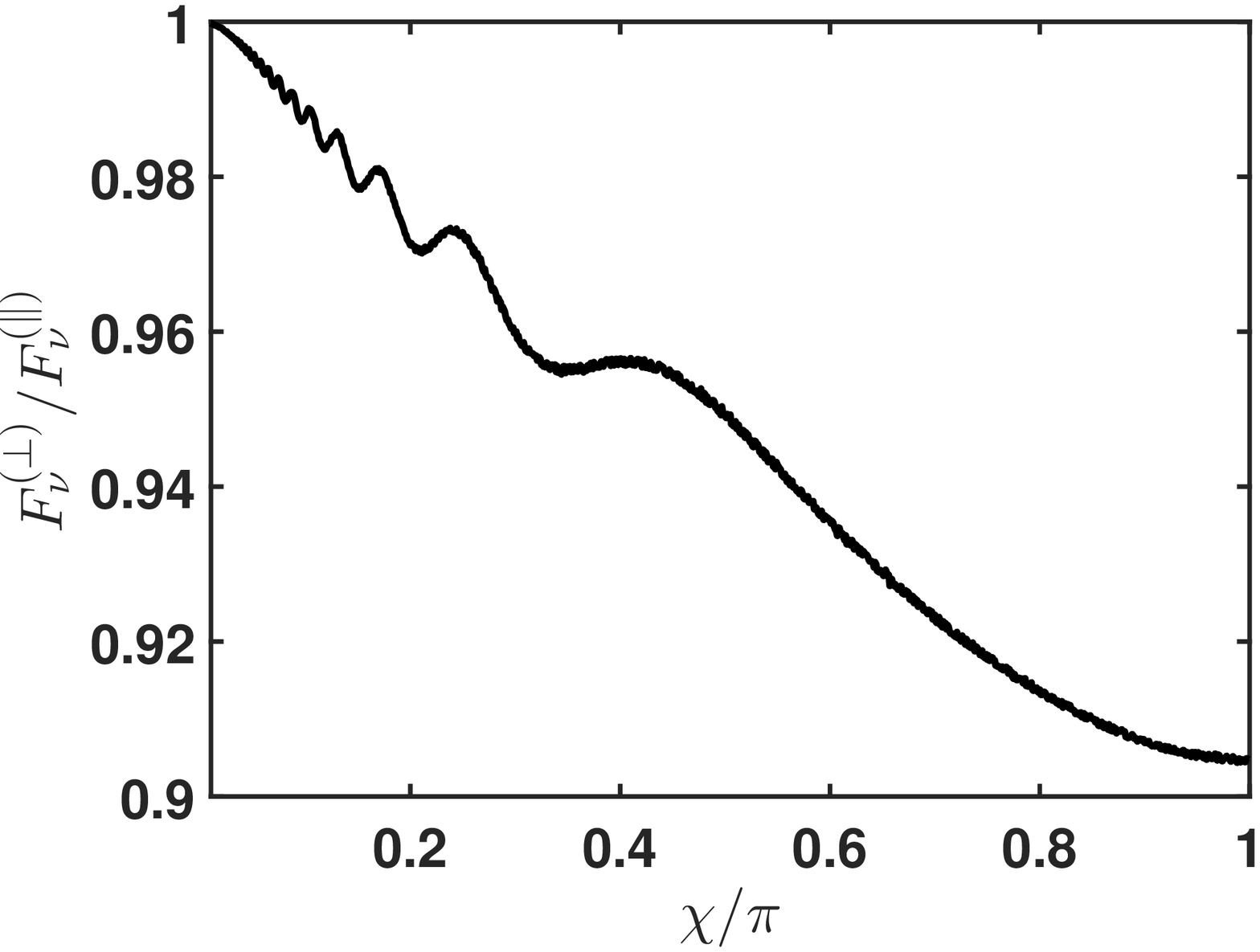}}
  \hskip-.6cm
  \subfigure[]
  {\label{5d}
  \includegraphics[scale=.35]{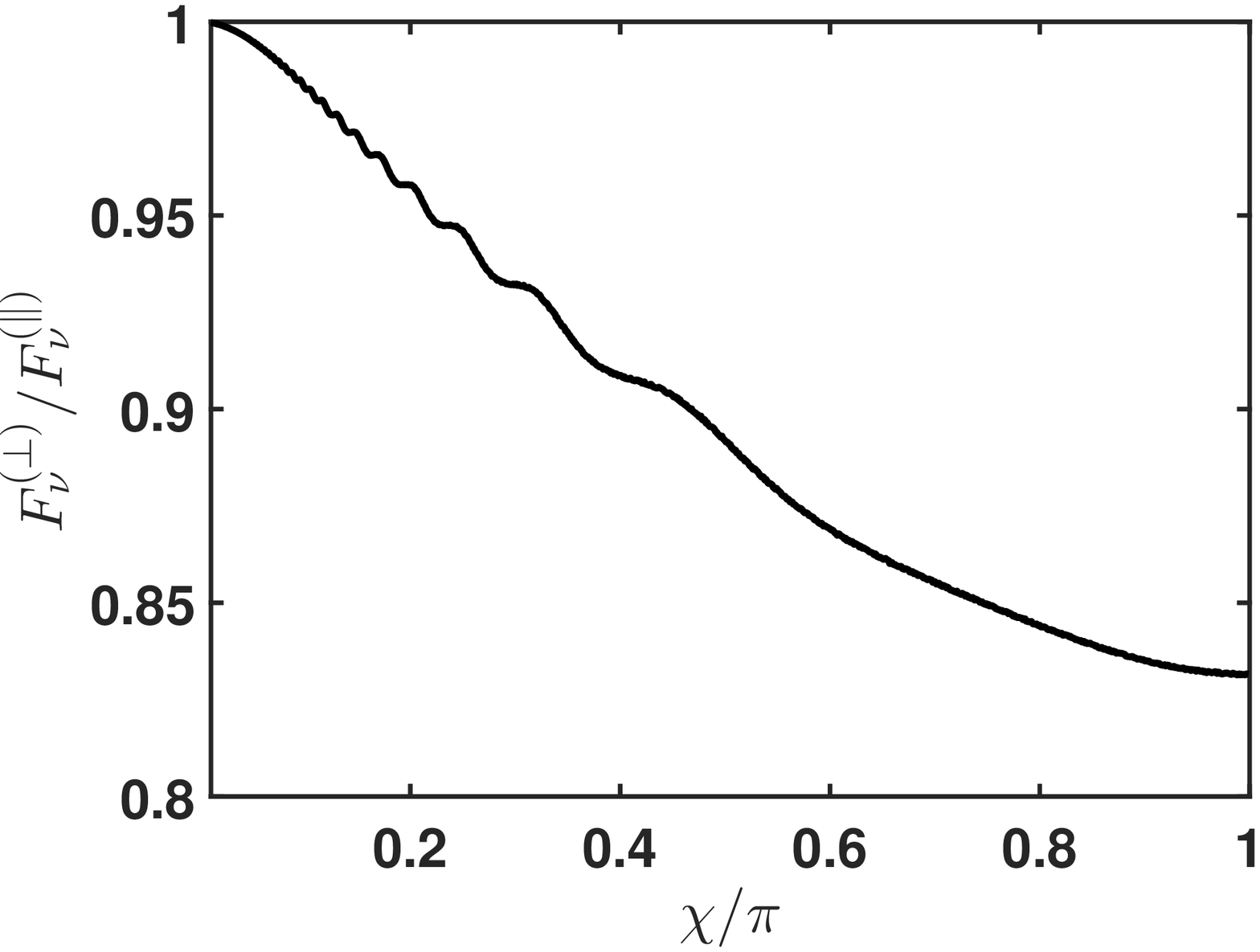}}
  \protect
  \caption{The ratio $F_{\nu}^{(\perp)}/F_{\nu}^{(\parallel)}$ versus the scattering
  angle $\chi$ for different $V_{\mathrm{max}}$ and $\beta$. (a)~$V_{\mathrm{max}}=0.1$
  ($n_{e}=10^{18}\,\text{cm}^{-3}$) and $\beta=0.5$; (b)~$V_{\mathrm{max}}=0.2$
  ($n_{e}=2\times10^{18}\,\text{cm}^{-3}$) and $\beta=0.5$; (c)~$V_{\mathrm{max}}=0.1$
  and $\beta=0.2$; (d)~$V_{\mathrm{max}}=0.2$ and $\beta=0.2$.\label{fig:ratflux}}
\end{figure}

Now we explicitly compare $\perp$ and $\parallel$ cases by plotting
the ratios of the corresponding fluxes in Fig.~\ref{fig:ratflux}.
First, we mention that $F_{\nu}^{(\perp)}<F_{\nu}^{(\parallel)}$.
Indeed, if a neutrino interacts only with gravity, spin oscillations
are in the resonance (see Refs.~\cite{Dvo06,Dvo13}). The interaction
with matter makes the survival probability greater. This fact explains
the observed feature.

The difference between $F_{\nu}^{(\perp)}$ and $F_{\nu}^{(\parallel)}$
can reach almost 20\%; see Fig.~\ref{5d}. It means that,
if high energy astrophysical neutrinos experience gravitational lensing
by BH surrounded by an accretion disk, the observed flux depends on
the orientation of the neutrino trajectory with respect to the disk
plane. This maximal difference between $F_{\nu}^{(\perp)}$ and $F_{\nu}^{(\parallel)}$
is for backwardly scattered neutrinos.

\section{Discussion\label{sec:DISC}}

In the present work, we have considered spin effects in the neutrino
scattering off a nonrotating BH. The neutrino spin evolution in curved
spacetime has been accounted for quasiclassically basing on the approach
developed in Refs.~\cite{Dvo06,Dvo13}. We have studied the neutrino
scattering off SMBH surrounded by an accretion disk and considered
some astrophysical applications.

In Sec.~\ref{sec:GRAV}, we have studied the neutrino spin evolution
in a gravitational scattering in the Schwarzschild metric. Supposing
that all incoming neutrinos are ultrarelativistic and left polarized,
we have obtained that the transition probability $P_{\mathrm{LR}}$
for outgoing particles can reach 25\% if the impact parameter is close
to the critical one $b\approx b_{0}=3\sqrt{3}r_{g}/2$. Note that
the fact that the helicity of ultrarelativistic (massless) particles
can be changed under the influence of a gravitational field was mentioned
earlier in Ref.~\cite{SinMobPap04}. We also mention that our calculation
of the probabilities in the limit $y\gg y_{0}$ in Eq.~(\ref{eq:Plim})
is consistent with the result of Ref.~\cite{Mer95}, where the neutrino
helicity flip in the idealized gravitational field was studied.

Then, in Sec.~\ref{sec:MATT}, we have derived the effective Schr\"{o}dinger
equation for a neutrino scattering off BH surrounded by background
matter with a nonuniform density. In the case of only gravitational
scattering, studied in Sec.~\ref{sec:GRAV}, it was possible to obtain
analytically the transition and survival probabilities for some impact
parameters. If, besides gravity, a neutrino interacts with a background
matter, the probabilities can be derived only in the numerical solution
of Eqs.~(\ref{eq:Schr}) and~(\ref{eq:Omegar}).

In Sec.~\ref{sec:APPL}, have considered the astrophysical applications
of our results. In particular, we have studied the effect of spin
oscillations on the neutrino scattering off SMBH surrounded by the
accretion disk. We have taken the parameters of the accretion disk,
such as the number density and the mass distribution, close to the
values resulting from observations and hydrodynamics simulations.
Using the numerical solution of Eqs.~(\ref{eq:Schr}) and~(\ref{eq:Omegar}),
we have found the transition and survival probabilities, as well as
the observed fluxes of outgoing neutrinos for different orientations
of the particles trajectories with respect to the accretion disk.

As one can see in Figs.~\ref{1c} and~\ref{fig:Fparal},
there is no deviation of the fluxes for the forward neutrino scattering
at $\chi=0$ if one compares them with the fluxes of scalar particles.
It means that neutrino spin oscillations do not affect the size of
the BH shadow. The major effect of spin oscillations is for the backward
neutrino scattering at $\chi=\pi$. Thus the intensity of the glory
flux for neutrinos is almost 20\% less than for scalar particles;
cf. Fig.~\ref{4a}.

The influence of the plasma interaction on the gravitational scattering
of scalar particles (photons) was extensively studied (see, e.g.,
Ref.~\cite{CunHer18} for a review). For example, the photons propagation
in plasma surrounding a nonrotating BH was examined in Ref.~\cite{PerTsu17}.
The form of the BH shadow was found to be unchanged, but its size
can be magnified. In Fig.~\ref{fig:ratflux}, we predict the asymmetry
in the observed neutrino fluxes depending on the orientation of the
neutrino trajectory with respect to the accretion disk. This asymmetry
is maximal for the backward neutrino scattering. Although neutrinos
interact with plasma much weaker than photons, the asymmetry can reach
almost 20\% for the realistic accretion disk; cf. Fig.~\ref{5d}.

Since the effects of spin oscillations on the neutrino gravitational
scattering are valid for ultrarelativistic particles, the results
obtained in the present work are of importance for the rapidly developing
area of the neutrino astronomy~\cite{Gal18}, where a significant
success was achieved in the detection of the ultrahigh energy (UHE)
cosmic neutrinos. Neutrinos with energies in the PeV range were reported
in Ref.~\cite{Aar13} to be detected. Moreover there are sizable
efforts in the identification of the sources of UHE neutrinos with
astronomical objects such as active galactic nuclei~\cite{Aar19}.
In our work we have demonstrated that, if the incoming flux of cosmic
neutrinos experience the gravitational lensing, the observed flux
can be reduced up to 20\%, compared to its initial value, because
of neutrino spin oscillations.

\section*{Acknowledgments}

I am thankful to J.~Jiang, Y.~N.~Obukhov, and A.~F.~Zakharov
for useful comments. This work is performed within the government
assignment of IZMIRAN. I am also thankful to RFBR (Grant No 18-02-00149a)
and DAAD for a partial support.

\appendix

\section{Particle motion in the Schwarzschild metric\label{sec:PARTM}}

In this Appendix, we briefly remind how to describe the motion of
a scalar particle interacting with a nonrotating BH, as well as how
to calculate the differential cross section of the gravitational scattering.
These problems were studied in details in Refs.~\cite[pp.~287-290]{LanLif71} and~\cite{DolDorLas06}.

The energy $E$ and the angular momentum $L$ are conserved quantities
for a particle with the mass $m$ moving in the Schwarzschild metric
in Eq.~(\ref{eq:intschw}). The equation of motion and the trajectory
are defined by
\begin{equation}
\frac{\mathrm{d}r}{\mathrm{d}t}=\pm\frac{mA^{2}}{E}\left[\frac{E^{2}}{m^{2}}-A^{2}\left(1+\frac{L^{2}}{m^{2}r^{2}}\right)\right]^{1/2},\quad\frac{\mathrm{d}\phi}{\mathrm{d}r}=\pm\frac{L}{mr^{2}}\left[\frac{E^{2}}{m^{2}}-A^{2}\left(1+\frac{L^{2}}{m^{2}r^{2}}\right)\right]^{-1/2},\label{eq:eqmtr}
\end{equation}
for a particle moving in the equatorial plane. In Eq.~(\ref{eq:eqmtr}),
the minus signs stay for incoming particles and the plus ones for
outgoing particles.

If we study ultrarelativistic particles, the angle corresponding to
the minimal distance between a particle and BH is
\begin{equation}
\phi_{m}=y\int_{x_{m}}^{\infty}\frac{dx}{\sqrt{x[x^{3}-y^{2}(x-1)]}},
\end{equation}
where $y=b/r_{g}$, $b=L/E$ is the impact parameter, and $x_{m}$
is the maximal root of Eq.~(\ref{eq:eqtosolve}). The parameter $y>y_{0}=3\sqrt{3}/2$.
Otherwise a particle falls to BH.

While computing the differential cross section, $\mathrm{d}\sigma/\mathrm{d}\varOmega$,
where $\mathrm{d}\varOmega=2\pi\sin\chi\mathrm{d}\chi$, we should
take into account that a particle, before being scattered off, can
make multiple revolutions around BH, both clockwisely and anticlockwisely.
One should account for this fact in the determination of the angle
$\chi$, fixing the position of a detector, which is in the range
$0<\chi<\pi$.

\begin{figure}
  \centering
  \includegraphics[scale=0.35]{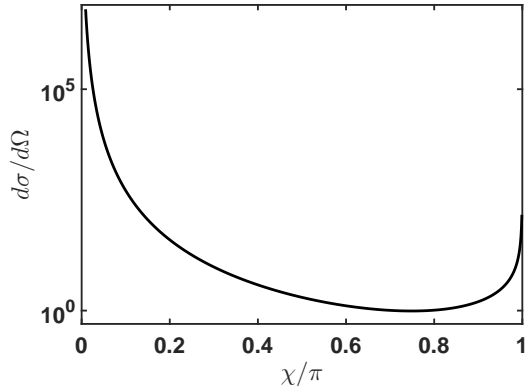}
  \caption{The differential cross section of the gravitational scattering of
  scalar particles off a nonrotating BH, normalized by $r_g^2$, versus the angle $\chi$.  
  \label{fig:diffsc0}}
\end{figure}

In Fig.~\ref{fig:diffsc0}, we present the result of the numerical
computation of the cross section. While building this plot, we take
that $y_{0}<y<30y_{0}$ and account for up to two revolutions of a
particle around BH in both directions. This our result is used in
Sec.~\ref{sec:APPL} when we study the neutrino scattering off a
realistic BH surrounded by an accretion disk.

\begin{thebibliography}{50}

\bibitem{Ago18}
  M.~Agostini \textit{et al.} (Borexino Collaboration),
  Comprehensive measurement of $pp$-chain solar neutrinos,
  Nature (London) \textbf{562}, 505\textendash 510 (2018).

\bibitem{GiuStu15}
  C.~Giunti and A.~Studenikin,
  Neutrino electromagnetic interactions: A window to new physics,
  Rev. Mod. Phys. \textbf{87}, 531\textendash 591 (2015)
  [arXiv:1403.6344].

\bibitem{MalSmi16}
  M.~Maltoni and A.~Yu.~Smirnov,
  Solar neutrinos and neutrino physics,
  Eur. Phys. J. A \textbf{52}, 87 (2016)
  [arXiv:1507.05287].

\bibitem{AhlBur96}
  D.~V.~Ahluwalia and C.~Burgard,
  Gravitationally induced neutrino-oscillation phases,
  Gen. Relativ. Gravit. \textbf{28}, 1161\textendash 1170 (1996)
  [gr-qc/9603008].

\bibitem{Dvo06}
  M.~Dvornikov,
  Neutrino spin oscillations in gravitational fields,
  Int. J. Mod. Phys. D \textbf{15}, 1017\textendash 1034 (2006)
  [hep-ph/0601095].

\bibitem{Dvo13}
  M.~Dvornikov,
  Neutrino spin oscillations in matter under the influence of gravitational
  and electromagnetic fields,
  J. Cosmol. Astropart. Phys. 06 (2013) 015
  [arXiv:1306.2659].

\bibitem{Dvo19}
  M.~Dvornikov,
  Neutrino spin oscillations in external fields in curved spacetime,
  Phys. Rev. D \textbf{99}, 116021 (2019)
  [arXiv:1902.11285].

\bibitem{ObuSilTer09}
  Yu.~N.~Obukhov, A.~J.~Silenko, and O.~V.~Teryaev,
  Spin dynamics in gravitational fields of rotating bodies
  and the equivalence principle,
  Phys. Rev. D \textbf{80}, 064044 (2009)
  [arXiv:0907.4367].

\bibitem{ObuSilTer17}
  Y.~N.~Obukhov, A.~J.~Silenko, and O.~V.~Teryaev,
  General treatment of quantum and classical spinning particles in external fields,
  Phys. Rev. D \textbf{96}, 105005 (2017)
  [arXiv:1708.05601].

\bibitem{SorZil07}
  F.~Sorge and S.~Zilio,
  Neutrino spin flip around a Schwarzschild black hole,
  Class. Quantum Grav. \textbf{24}, 2653\textendash 2664 (2007).

\bibitem{AlaNod15}
  S.~A.~Alavi and S.~Nodeh,
  Neutrino spin oscillations in gravitational fields in noncommutative spaces,
  Phys. Scripta \textbf{90}, 035301 (2015)
  [arXiv:1301.5977].

\bibitem{Cha15}
  S.~Chakraborty,
  Aspects of neutrino oscillation in alternative gravity theories,
  J. Cosmol. Astropart. Phys. 10 (2015) 019
  [arXiv:1506.02647].

\bibitem{Aki19}
  K.~Akiyama \textit{et al.} (Event Horizon Telescope Collaboration),
  First M87 event horizon telescope results.
  I. The shadow of the supermassive black hole,
  Astrophys. J. Lett. \textbf{875}, L1 (2019).

\bibitem{GraHolWal19}
  S.~E.~Gralla, D.~E.~Holz, and R.~M.~Wald,
  Black hole shadows, photon rings, and lensing rings,
  Phys. Rev. D \textbf{100}, 024018 (2019)
  [arXiv:1906.00873].

\bibitem{CabMcLSur12}
  O.~L.~Caballero, G.~C.~McLaughlin, and R.~Surman,
  Neutrino spectra from accretion disks:
  Neutrino general relativistic effects and the consequences for nucleosynthesis,
  Astrophys. J. \textbf{745}, 170 (2012)
  [arXiv:1105.6371].

\bibitem{Cor15}
  C.~Corian\`o, A.~Costantini, M.~Dell'Atti, and L.~Delle Rose,
  Neutrino and photon lensing by black holes:
  radiative lens equations and post-Newtonian contributions,
  J. High Energy Phys. 07 (2015) 160
  [arXiv:1504.01322].

\bibitem{StuSch19}
  Z.~Stuchl\'{i}k and J.~Schee,
  Shadow of the regular Bardeen black holes and comparison of the motion of photons
  and neutrinos,
  Eur. Phys. J. C \textbf{79}, 44 (2019).

\bibitem{CunHer18}
  P.~V.~P.~Cunha and C.~A.~R.~Herdeiro,
  Shadows and strong gravitational lensing: a brief review,
  Gen. Relativ. Gravit. \textbf{50}, 42 (2018)
  [arXiv:1801.00860].

\bibitem{LanLif71}
  L.~D.~Landau and E.~M.~Lifschitz,
  \textit{The Classical Theory of Fields}, 3rd ed.
  (Pergamon Press, Oxford, 1971).

\bibitem{MohPal04}
  R.~N.~Mohapatra and P.~B.~Pal,
  \textit{Massive Neutrinos in Physics and Astrophysics}, 3rd ed.
  (World Scientific, Singapore, 2004), p.~98.

\bibitem{DvoStu02}
  M.~Dvornikov and A.~Studenikin,
  Neutrino spin evolution in presence of general external fields,
  J. High Energy Phys. 09 (2002) 016
  [hep-ph/0202113].

\bibitem{Igu00}
  I.~V.~Igumenshchev, M.~A.~Abramowicz, and R.~Narayan,
  Numerical simulations of convective accretion flows in three dimensions,
  Astrophys. J. \textbf{537}, L27\textendash L30 (2000).

\bibitem{Jia19}
  J.~Jiang, A.~C.~Fabian, T.~Dauser, L.~Gallo, J.~A.~Garcia, E.~Kara, M.~L.~Parker, 
  J.~A.~Tomsick, D.~J.~Walton, and C.~S.~Reynolds,
  High Density Reflection Spectroscopy \textendash{}
  II. The density of the inner black hole accretion disc in AGN,
  Mon. Not. R. Astron. Soc. \textbf{489}, 3436\textendash 3455 (2019)
  [arXiv:1908.07272].

\bibitem{DolDorLas06}
  S.~Dolan, C.~Doran, and A.~Lasenby,
  Fermion scattering by a Schwarzschild black hole,
  Phys. Rev. D \textbf{74}, 064005 (2006)
  [gr-qc/0605031].

\bibitem{SinMobPap04}
  D.~Singh, N.~Mobed, and G.~Papini,
  Helicity precession of spin-1/2 particles in weak inertial and gravitational fields,
  J. Phys. A: Math. Gen. \textbf{37}, 8329\textendash 8347 (2004)
  [hep-ph/0405296].

\bibitem{Mer95}
  C.~Mergulh\~{a}o Jr.,
  Neutrino helicity flip in a curved space-time,
  Gen. Relativ. Gravit. \textbf{27}, 657\textendash 667 (1995).

\bibitem{PerTsu17}
  V.~Perlick and O.~Yu.~Tsupko,
  Light propagation in a plasma on Kerr spacetime:
  Separation of the Hamilton-Jacobi equation and calculation of the shadow,
  Phys. Rev. D \textbf{95}, 104003 (2017)
  [arXiv:1702.08768].

\bibitem{Gal18}
  A.~Gallo Rosso, C.~Mascaretti, A.~Palladino, and F.~Vissani,
  Introduction to neutrino astronomy,
  Eur. Phys. J. Plus \textbf{133}, 267 (2018)
  [arXiv:1806.06339].

\bibitem{Aar13}
  M.~G.~Aartsen \textit{et al.} (IceCube Collaboration),
  First observation of PeV-energy neutrinos with IceCube,
  Phys. Rev. Lett. \textbf{111}, 021103 (2013)
  [arXiv:1304.5356].

\bibitem{Aar19}
  M.~G.~Aartsen \textit{et al.} (IceCube Collaboration),
  Time-integrated neutrino source searches with 10 years of IceCube data,
  submitted to Phys. Rev. Lett.
  [arXiv:1910.08488].

\end{thebibliography}
\end{document}